\documentstyle[aps,prb]{revtex}
\newcommand\prr[1]{\protect\ref{#1}\protect}
\newcommand\prcite[1]{\protect\cite{#1}\protect}
\newcommand\sla{{\sl a}}
\newcommand\slb{{\sl b}}
\newcommand\figmacro[1]{\begin{figure}\caption{#1}\end{figure}}
\begin{document}
\begin{center}
{\it Semiconductor Science and Technology} \\[2cm]
{\bf Review article on:}
\end{center}
\bigskip
\title{Coherent phenomena in semiconductors}

\author{Fausto Rossi}
\address{
Istituto Nazionale per la Fisica della Materia (INFM) and \\
Dipartimento di Fisica, Universit\`a di Modena \\ 
via Campi 213/A, I-41100 Modena, Italy \\
E-mail:\ \ Rossi@UNIMO.IT
}
\begin{abstract}

A review of coherent phenomena in photoexcited semiconductors is presented.
In particular, two classes of phenomena are considered: On the one hand 
the role played by optically-induced phase coherence in the ultrafast 
spectroscopy of semiconductors; On the other hand 
the Coulomb-induced effects on the coherent optical response of low-dimensional
structures.

All the phenomena discussed in the paper are analyzed in terms of a 
theoretical framework based on the density-matrix formalism. 
Due to its generality, this quantum-kinetic approach allows a realistic 
description of coherent as well as incoherent, i.e. phase-breaking, 
processes, thus providing quantitative information on the coupled 
---coherent vs. incoherent--- carrier dynamics in photoexcited 
semiconductors.

The primary goal of the paper is to discuss the concept of 
quantum-mechanical phase 
coherence as well as its relevance and implications on semiconductor physics 
and technology.
In particular, we will discuss the dominant role played by optically 
induced phase coherence on the process of carrier photogeneration and 
relaxation in bulk systems. 
We will then review typical field-induced 
coherent phenomena in semiconductor superlattices such as Bloch oscillations and
Wannier-Stark localization. 
Finally, we will discuss the dominant role played by Coulomb 
correlation on the linear and non-linear optical spectra of realistic 
quantum-wire structures.

\end{abstract}
\maketitle
\section{Introduction}\label{s.int}

The resonant excitation by coherent optical radiation of an electronic 
transition in a semiconductor, e.g. a valence- to conduction-band excitation, 
creates a quantum-mechanical coherent superposition of the initial and 
final states of the transition, called optical polarization. 
The non-linear optical 
properties of this coherent superposition, together with its time 
evolution,
can be used to provide a sensitive 
measurement of many fundamental parameters of the semiconductor material 
\prcite{Shah96} including elastic 
and inelastic scattering, energy level splittings between nearly degenerate
states, energy relaxation, as well as associated information such as Land\'e
g-factors and band-mixing. 

In general, non-linear laser spectroscopy is a very established and well 
understood means to probe these phenomena 
\prcite{Allen75,Sargent77,Shen84,Levenson88,Meystre91,Mills91,Haug93,Demtroder96}. 
However, the description of the non-linear optical response of a 
semiconductor crystal can be considerably 
more complex than for simple isolated and non-interacting atoms \prcite{Haug93}. 
This is particularly true for the case of the ultrafast optical 
spectroscopy used for the study of the sub-picosecond dynamics 
of photoexcited carriers in bulk systems as well as in semiconductor 
heterostructures \prcite{Shah96,Henneberger93,Phillips94}.

The life-time of the coherent quantum-mechanical superposition 
generated by an ultrafast laser excitation, called ``dephasing time'', 
determines the typical time-scale on which coherent phenomena can be 
observed. Such dephasing time reflects the role played by the various 
``incoherent'', i.e. phase-breaking, mechanisms in destroying the phase 
coherence induced by the laser photoexcitation.
Since semiconductors are characterized by very short electron-hole
dephasing times, 
of the order of few hundreds of femtoseconds \prcite{Haug93}, coherent 
phenomena manifest themselves only through ultrafast optical experiments 
with sub-picosecond time resolutions \prcite{Shah96}.
On this time-scale, the ultrafast evolution of photoexcited electron-hole 
pairs will reflect the strong coupling between coherent and incoherent 
dynamics, thus providing invaluable information on the non-equilibrium 
relaxation processes occurring in the semiconductor,
e.g. carrier-carrier and carrier-phonon scattering.

The aim of this paper is to review the basic aspects related to coherent 
phenomena in semiconductors. In particular, we will focus on 
the ultrafast ---coherent vs. incoherent---
carrier dynamics as well as on  
Coulomb-correlation effects 
in photoexcited semiconductors. 

The paper is organized as follows. 
In the remainder of this section, after a brief historical account of 
coherent experiments in solids, we will try to gain more insight into the 
concept of coherence by introducing a simplified description of the 
light-matter interaction in terms of a two-level model.
In section \prr{s.tb} we will discuss the theoretical 
approach commonly used for a realistic description of both coherent and 
incoherent phenomena in various semiconductor structures, e.g. bulk 
systems, semiconductor superlattices, quantum wells and wires.
Section \prr{s.bulk} deals with ultrafast carrier photoexcitation and 
relaxation in bulk semiconductors; In particular, we will discuss the 
dominant role played by coherence on the 
carrier photogeneration process.
In section \prr{s.sl} we will review and discuss typical field-induced 
phenomena in superlattices, namely Bloch 
oscillations and Wannier-Stark localization, as well as their dephasing 
dynamics. 
Section \prr{s.qwr} is devoted to the analysis of the coherent optical 
response of quasi-one-dimensional systems; More specifically, we will 
discuss the strong modifications induced by Coulomb correlation on the 
linear and non-linear optical spectra of realistic quantum-wire structures.
Finally, in section \prr{s.suco} we will summarize and draw some conclusions.

\subsection{Historical background}\label{ss.hb}

Coherent phenomena in atomic and molecular systems have been investigated 
for a long time \prcite{Allen75}. 
The first spin echo experiment \prcite{Hahn50} 
was performed in 1950 on protons in a water solution of Fe$^{+++}$ 
ions. 
Pulses in the radio frequency range were generated by means of a gated 
oscillator with pulse widths between 20~$\mu$s and a few milliseconds. 
With these pulses 
dephasing times of the order of 10 ms have been measured.

In the 1960s echo experiments were brought into the visible 
range \prcite{Kurnit64,Abella66}.
A Q-switched ruby laser produced pulses of 
approximately 10 ns duration which were used to observe photon echoes from 
ruby. In this case the dephasing times were of the order of 100 ns.

As already pointed out, for the observation of any coherent dynamics 
the pulse width has to be shorter than the typical dephasing time. 
Since in semiconductors electron-hole dephasing times are much shorter 
---they are in the range of a few picoseconds down to some femtoseconds---,
coherent experiments in semiconductors had to wait until the development of 
suitable lasers able to generate sub-picosecond pulses. 

\subsection{Ultrafast spectroscopy in semiconductors}\label{ss.uss}

The physical phenomena governing the ultrafast carrier dynamics in
photoexcited semiconductors can be divided into two classes: 
{\it coherent phenomena}, i.e. phenomena related to the 
quantum-mechanical phase coherence induced by the laser photoexcitation, 
and {\it incoherent phenomena}, i.e. phenomena induced by the 
various phase-breaking scattering mechanisms.
The above classification in terms of coherent and incoherent phenomena is 
not purely academic; It corresponds to rather different experimental 
techniques for the investigation of these two different regimes.

From an historical point of view, the optical spectroscopy in 
semiconductors started with the analysis of relatively slow 
incoherent phenomena (compared to the electron-hole dephasing time-scale).
The investigation of nonequilibrium carriers started with the analysis 
of the incoherent energy-relaxation processes 
in the late 1960s using cw-lasers \prcite{Shah69}. 
In the 1970s pulse sources for the study of photoexcited carriers 
became available \prcite{Shank79} and this initiated 
time-resolved studies of the energy-relaxation process. 
Many experiments based on different techniques have been 
performed \prcite{Shah96}: 
band-to-band luminescence 
\prcite{Shah87,Elsaesser91} which monitors the product of 
electron and hole distribution functions, 
band-to-acceptor luminescence \prcite{Ulbrich89,Peterson90,Snoke92} which 
provides information on the electron distribution functions only, 
and pump-and-probe measurements 
\prcite{Oudar84,Foing92,Leitenstorfer96a,Lutgen96,Joschko97} 
where the measured differential transmission is proportional to the sum of 
electron and hole distribution functions. 

The theoretical analysis of these relaxation phenomena is commonly  
based on the semiclassical Boltzmann theory. 
The Boltzmann equations for both electron and hole distribution functions are 
commonly solved by means of semiclassical Monte Carlo simulations 
\prcite{Jacoboni89,Goodnick92,Rossi92a}.

In addition to the analysis of incoherent energy-relaxation processes,
the ultrafast optical spectroscopy has allowed the investigation of 
coherent phenomena. 
As already pointed out, a coherent laser field creates, in addition to a 
non-equilibrium carrier distribution, a coherent polarization. 
The investigation 
of the coherent dynamics in semiconductors 
started in the 1980s. Different aspects have been 
investigated \prcite{Shah96}: 
the optical Stark effect \prcite{Mysyrowicz86,Peyghambarian89,Knox89,Cundiff94}, 
the dephasing of
free carriers \prcite{Becker88} and excitons 
\prcite{Masumoto83,Schultheis86,Noll90},
quantum beats related to various types of level splittings 
\prcite{Langer90,Gobel90,Frohlich91,Stolz91}, 
charge oscillations in double-quantum-well systems 
\prcite{Leo91,Feldmann93} and superlattices \prcite{Waschke93},
many-particle effects \prcite{Leo90,Weiss92,Kim92a,Kim92b},
and the emission of coherent THz radiation \prcite{Roskos92,Planken92}.

These experiments cannot be analyzed within the framework of 
the Boltzmann equation. The reason is that this coherent polarization 
reflects a well-defined phase relation between electrons and holes, 
which is neglected within the semiclassical Boltzmann theory. 
Any proper description requires a quantum-mechanical treatment where, 
in addition to the 
distribution functions of electrons and holes, also
the interband polarization is taken into account as an independent
variable. Several approaches have been used in the 
literature \prcite{Haug93}: 
Bogoliubov transformations \prcite{Comte86},
nonequilibrium Green's functions 
\prcite{Schmitt86a,Schmitt88,Henneberger88,Kuznetsov91,Haug92},
band-edge equations based on the real-space density matrix
\prcite{Balslev89},
and the density matrix formalism in momentum space
\prcite{Schmitt86b,Lindberg88,Wegener90,Schafer93,Binder94,Hess96,Kuhn97}.

During the last decade the time resolution has been further improved down 
to few tens of femtoseconds \prcite{Fork87}.
On such extremely short time-scale coherent effects can no longer be 
neglected and are found to play a dominant role 
also for the case of typical incoherent measurements such as 
time-resolved and time-integrated luminescence. 
In such conditions, the carrier dynamics is the result of a strong 
interplay between coherent and incoherent phenomena. Therefore, the 
traditional separation 
\footnote{
From an historical point of view, the theoretical approaches commonly used 
for the interpretation of coherent phenomena provide a rather qualitative 
description of phase-breaking processes in terms of phenomenological 
energy-relaxation and dephasing rates. On the other hand, incoherent 
phenomena were traditionally investigated in terms of semiclassical 
Monte Carlo simulations, which provide a microscopic description of the 
non-equilibrium scattering dynamics.
}
between coherent and incoherent approaches for 
the theoretical investigation of photoexcited semiconductors is no longer 
valid. What is needed is a comprehensive theoretical framework 
able to describe on the same kinetic level both classes of phenomena 
as well as their mutual coupling. 
To this purpose, a generalized Monte Carlo method for the analysis of both 
coherent and incoherent phenomena has been recently proposed 
\prcite{Kuhn92a,Kuhn92b,Rossi93,Rossi94,Haas96}.
The spirit of the method is to combine the advantages of the 
conventional Monte Carlo approach \prcite{Jacoboni89,Goodnick92,Rossi92a} 
in treating the incoherent, i.e. phase-breaking, dynamics with the strength of 
a quantum-kinetic approach in describing coherent phenomena \prcite{Haug93}. 
In particular, the coherent contributions are evaluated by means of a direct 
numerical integration while the incoherent ones are ``sampled'' by means of a
conventional Monte Carlo simulation in the three-dimensional 
${\bf k}$-space.

This theoretical approach has ben applied 
successfully to the analysis of various ultrafast optical experiments, e.g. 
four-wave-mixing studies of many-body effects in bulk GaAs 
\prcite{Lohner93,Leitenstorfer94a} and 
luminescence studies of the hot-carrier photogeneration process 
\prcite{Leitenstorfer94b,Kuhn95,Leitenstorfer96b}.

\subsection{The meaning of coherence}\label{ss.moc}

In order to clarify the concept of phase coherence, 
let us consider the simplest physical model for the description of 
light-matter interaction, i.e. an optically driven two-level system 
\prcite{Shah96,Allen75}.
Within a two-level picture, a ground state $\sla$ with energy 
$\epsilon_\sla$ and an excited state $\slb$ with energy 
$\epsilon_\slb$ are mutually coupled by a driving force, e.g. an external 
field, and/or by their mutual interaction, e.g. Coulomb correlation. 
The two-level-system Hamiltonian 
\begin{equation}\label{eq1.1}
{\bf H} = {\bf H}_\circ + {\bf H}'
\end{equation}
is the sum of a free-level contribution
\begin{equation}\label{eq1.2}
{\bf H}_\circ = \epsilon_\sla {a}^\dagger_\sla {a}^{ }_\sla +
\epsilon_\slb {a}^\dagger_\slb {a}^{ }_\slb
\end{equation}
and of a coupling term
\begin{equation}\label{eq1.3}
{\bf H}' = 
U_{\slb\sla} {a}^\dagger_\slb {a}^{ }_\sla +
U_{\sla\slb} {a}^\dagger_\sla {a}^{ }_\slb\ .
\end{equation}
Here, the usual second-quantization picture in terms of creation 
(${a}^\dagger$) and destruction (${a}$) operators has been introduced 
\prcite{Haug93}.
The two terms forming the coupling Hamiltonian ${\bf H}'$ 
will induce transitions 
from state $\sla$ to $\slb$ and vice versa according to the coupling constant 
$U^{ }_{\slb\sla} = U^*_{\sla\slb}$.

As a starting point, let us consider a single-electron system.
In the absence of interlevel coupling ($U_{\sla\slb} = 0$), we have 
two stationary states,
\begin{equation}\label{eq1.4}
\vert \sla(t)\rangle = exp\left({-{i\epsilon_\sla t\over \hbar}} \right)
{a}^\dagger_\sla \vert 0 \rangle\ , \qquad
\vert \slb(t)\rangle = exp\left({-{i\epsilon_\slb t\over \hbar}}\right)
{a}^\dagger_\slb \vert 0 \rangle
\end{equation}
($\vert 0 \rangle$ being the vacuum state), 
corresponding to a single electron in level $\sla$ or $\slb$, respectively.
On the contrary,
in the presence of interlevel coupling, the state of the system 
is, in general, a linear superposition of the 
non-interacting states in equation (\prr{eq1.4}):
\begin{equation}\label{eq1.5}
\vert\psi(t)\rangle = c_\sla(t) \vert \sla(t) \rangle + 
c_\slb(t) \vert \slb(t) \rangle
\ ,
\end{equation}
whose coefficients obey the following equations of motion:
\begin{eqnarray}\label{eq1.6}
{d \over dt} c_\sla = && {U_{\sla\slb}\over i\hbar} 
exp\left({i(\epsilon_\sla-\epsilon_\slb) t \over\hbar}\right) c_\slb 
\nonumber \\
{d \over dt} c_\slb = && {U_{\slb\sla}\over i\hbar} 
exp\left({i(\epsilon_\slb-\epsilon_\sla) t \over\hbar}\right) c_\sla\ .
\end{eqnarray}
Again, we see that in the absence of interlevel coupling ($U = 0$) 
there is no time variation of the coefficients, 
i.e. if the system is prepared in state $\vert\sla\rangle$ 
or $\vert\slb\rangle$ it will remain in such eigenstate.
On the contrary, the interlevel coupling induces a time variation of 
the coefficients.

The above two-level model provides the simplest description of light-matter
interaction in atomic and molecular systems \prcite{Allen75} as well as in 
solids \prcite{Haug93}. 
The coupling term in equation (\prr{eq1.3}) has the same structure 
of a field-dipole interaction Hamiltonian and the driving force $u$ can be 
regarded as a coherent light field.
We will now consider two limiting cases: the ultrashort- and 
continuous-excitation regimes.

An ultrashort optical excitation can be described in terms of a delta-like 
light pulse: $U_{\sla\slb}(t) = \eta\delta(t)$. 
In this case, the equations of motion (\prr{eq1.6}) can be solved 
analytically: 
Due to this excitation, the two-level system will undergo an 
instantaneous transition from its ground state $\{c_\sla = 1, c_\slb = 0\}$ 
to the excited state $\{c_\sla = \cos\alpha, c_\slb = -i\sin\alpha\}$, 
with $\alpha = {\eta\over\hbar}$. 
Therefore, after the pulse the system will remain in the excited state
\begin{equation}\label{eq1.7}
\vert\psi(t)\rangle = 
(\cos\alpha) \vert \sla(t) \rangle + 
(-i\sin\alpha) \vert \slb(t) \rangle
\ ,
\end{equation}
which is a coherent quantum-mechanical superposition of the two 
non-interacting states. In addition to a finite occupation probability 
$|c_\slb|^2 = sin^2\alpha$ of the excited state, there exists a well 
defined phase coherence between the ground- and the excited-state 
contributions, i.e. a part from their amplitudes, 
the coefficients $c_\sla$ and $c_\slb$ differ in phase by ${\pi\over 2}$.
This is what is generally meant by optically induced phase coherence.

Let us now consider a continuous optical excitation resonant with
our two-level system: $U_{\sla\slb}(t) = U_\circ e^{i\omega_L t}$ with 
$\hbar\omega_L = \epsilon_\slb-\epsilon_\sla$.
Also for this case, equations (\prr{eq1.6}) can be solved 
analytically. 
In particular, for the initial condition at $t = 0$ given by the 
ground state $\{c_\sla = 1, c_\slb = 0\}$, the solution is given by 
\begin{equation}\label{eq1.8}
\vert\psi(t)\rangle = 
\cos({1 \over 2}\omega_R t) \vert \sla(t) \rangle - 
i\sin({1 \over 2}\omega_R t) \vert \slb(t) \rangle
\ ,
\end{equation}
where $\omega_R = {2U_\circ\over\hbar}$ is the so-called Rabi frequency. 
Compared to the previous case, the continuous excitation gives rise to a 
periodic population and depopulation of the excited state according to 
$|c_\slb|^2 = sin^2({1 \over 2}\omega_R t)$. This purely coherent phenomenon 
is known as ``Rabi-oscillation regime''.
As for the previous case, the excited state in equation (\prr{eq1.8}) 
reflects a well-defined optically-induced phase coherence. 

As mentioned above, the interlevel-coupling Hamiltonian in equation 
(\prr{eq1.3}) may also describe, in addition to an optical excitation, 
some interlevel coupling due, e.g. to Coulomb interaction.
Also in this case, in the presence of this interlevel coupling 
the quantum state of the system will result in a coherent superposition 
of the two stationary states in (\prr{eq1.4}).
Again, this coupling will induce a well-defined phase coherence 
between the two levels.

From the above considerations we see that coherent phenomena can be divided
into two basic classes: 
{\it optically-induced} and {\it Coulomb-induced} 
phenomena.
In order to describe the optical response of an electron gas within a 
semiconductor crystal, the above analysis, based on a single two-level 
system, has to be replaced by a statistical-ensemble approach, i.e. a 
description based on a collection of independent two-level systems.
Any statistical ensemble of quantum systems is properly described by its 
single-particle density matrix \prcite{Kuhn97}
 \begin{equation}\label{eq1.9}
\rho_{nn'} = \langle a^\dagger_{n'} a^{ }_n \rangle\ ,
\end{equation}
where $n$ denotes a generic set of quantum numbers. Its diagonal elements 
$f_n = \rho_{nn}$ provide the average occupation numbers while
the non-diagonal terms describe the degree of phase coherence 
between states $n$ and $n'$.
For the case of our two-level system, $\rho_{nn'}$ reduces to a 
two-by-two matrix: the diagonal elements $f_\sla = \rho_{\sla\sla}$ and 
$f_\slb = \rho_{\slb\slb}$ describe, respectively, the average occupation 
of levels $\sla$
and $\slb$, while the non-diagonal terms $p = 
\rho_{\slb\sla}$ and $p^* = \rho_{\sla\slb}$ reflect the average degree of 
phase coherence between the ground and the excited state.
Starting from the Heisenberg equations of motion for the creation and 
destruction operators, i.e.  
\begin{equation}\label{eq1.10}
i\hbar {d\over dt} a^\dagger_n = [a^\dagger_n,{\bf H}]
\ , \qquad
i\hbar {d\over dt} a^{ }_n = [a^{ }_n,{\bf H}]\ ,
\end{equation}
we obtain the following equations of motion for the above density-matrix 
elements:
\begin{eqnarray}\label{eq1.11}
{d \over dt} f_\slb = &&
-{d \over dt} f_\sla = {2\over\hbar}\Re\left(i U_{\sla\slb} p\right) 
\nonumber \\
{d \over dt} p = && {\epsilon_\slb-\epsilon_\sla\over i\hbar} p 
+{U_{\slb\sla}\over i\hbar} (f_\sla-f_\slb) \ .
\end{eqnarray}
They are known as optical Bloch equations \prcite{Shah96,Haug93} 
in analogy with the equations 
first derived by Bloch \prcite{Bloch46} for the spin systems.
This statistical-ensemble description reduces to the previous one 
for the case of a so called ``pure state'', i.e. the case in which all the 
two-level systems forming the ensemble are in the same quantum-mechanical state 
$\vert\psi\rangle$. 
In this case, the above set of kinetic equations and the equations of 
motion (\prr{eq1.6}) are totally equivalent. However, 
as we will discuss in section \prr{s.tb}, the density-matrix approach
introduced so far allows, in addition to the study of coherent phenomena, 
the analysis of incoherent phenomena, which is not possible within a 
simple Schr\"odinger-equation formalism.

The optical Bloch equations (\prr{eq1.10}) provide the simplest description 
of light-matter interaction. 
Let us consider again the case of a continuous optical excitation resonant 
with our two-level system: 
$U_{\sla\slb}(t) = U_\circ e^{i\omega_L t}$.
If we choose as initial condition at time $t = 0$ the state 
$\{f_\sla = f_\circ, f_\slb = 0, p = 0\}$, the solution of 
the above optical Bloch 
equations is given by:
\begin{eqnarray}\label{eq1.12}
f_\sla(t) = && f_\circ \cos^2({1 \over 2}\omega_R t) \nonumber \\
f_\slb(t) = && f_\circ \sin^2({1 \over 2}\omega_R t) \nonumber \\
p(t) = && {f_\circ\over 4} 
\left[
e^{-i\omega^+ t} -
e^{-i\omega^- t}\right]\ ,
\end{eqnarray}
where $\omega^\pm = \omega_L \pm \omega_R$. As for the case of a single 
two-level system, the above solution describes a Rabi-oscillation regime. 
In particular, the interlevel density-matrix element $p$ originates from 
the superposition of the two frequency components $\omega^+$ and $\omega^-$. 
They differ from $\omega_L$ by the Rabi frequency 
$\omega_R$. Such modification of the two-level frequency $\omega_L$ due to 
its coupling with the external field is known as ``Rabi splitting'' 
\prcite{Meystre91}.

If we now rewrite the interlevel density-matrix element $p$ 
in (\prr{eq1.12}) as
\begin{equation}\label{eq1.13}
p(t) = -{i f_\circ \over 2} e^{-i\omega_L t} \sin(\omega_R t)\ ,
\end{equation}
we see that, a part from the quantum-mechanical phase factor corresponding 
to the energy separation $\hbar\omega_L = \epsilon_\slb-\epsilon_\sla$, 
its amplitude exhibits Rabi oscillations 
according to $\sin(\omega_R t)$. More specifically, we obtain:
\begin{equation}\label{eq1.14}
|p|^2 \propto \sin^2(\omega_R t) \propto 
\cos^2({1 \over 2}\omega_R t) \sin^2({1 \over 2}\omega_R t) \propto
f_\sla(t) f_\slb(t)\ ,
\end{equation}
i.e. the quantity $|p|^2$ is proportional to the product of the two 
occupation numbers, thus reflecting the total (or macroscopic) 
dipole moment of the two-level system at time $t$.
This elucidates the link between optically-induced phase coherence and 
polarization: a coherent optical excitation gives rise to a 
coherent quantum-mechanical superposition of the two states which 
results in a macroscopic polarization of the system. Such polarization 
field is fully described by the non-diagonal matrix element $p$ in 
(\prr{eq1.12}). 

The above simplified description of light-matter interaction neglects any 
kind of incoherent, i.e. phase-breaking, phenomena. As we will see in section 
\prr{s.tb}, such dephasing processes will lead to a decay of the interband 
polarization $p$, thus destroying the optically induced phase 
coherence in the carrier system. 

As a final remark, let us discuss the concept of phase coherence in 
connection with the choice of representation.
On the basis of the density-matrix formalism considered so far, the phase 
coherence is described by the non-diagonal density-matrix 
elements $\rho_{nn'}$ in (\prr{eq1.9}). 
However, this separation in {\it diagonal} and {\it non-diagonal} terms is 
clearly representation dependent: what is diagonal in a given basis is
in general not diagonal in a different basis and vice versa. 
If one considers, as basis set, the eigenstates of the total Hamiltonian 
${\bf H}$ (which includes the external driving force), the density matrix 
$\rho$ in (\prr{eq1.9}) is always diagonal, i.e. no phase coherence.
Thus, in order to speak of phase coherence, we need to regard the total 
Hamiltonian ${\bf H}$ as the sum of the system Hamiltonian ${\bf H}_\circ$ 
plus a driving-force term ${\bf H}'$ (see equation (\prr{eq1.1})). 
Provided such separation between {\it system of interest} and 
{\it driving force}, the 
non-diagonal density-matrix elements 
within the representation given by the eigenstates of 
${\bf H}_\circ$ will describe the degree of coherence induced in the system by 
the driving force ${\bf H}'$. 

Two typical cases, discussed respectively in sections \prr{s.sl}
and \prr{s.qwr}, may help in clarifying the meaning of coherence.
The first case is that of a semiconductor superlattice in the presence of 
a constant and homogeneous electric field: As we will see, within the 
Wannier-Stark representation the Bloch-oscillation dynamics is 
the result of phase coherence between the various Wannier-Stark states; 
On the contrary, within the Bloch representation the 
same phenomenon is purely described in terms of carrier populations, i.e. 
no off-diagonal density-matrix elements.
The second case is that of Coulomb-induced phase coherence: Within a 
free-particle representation, even at the simplest Hartree-Fock level, 
Coulomb interaction induces phase coherence 
between free-carrier states.
This is clearly not the case within an exciton basis, where the density matrix 
is diagonal.

\section{Theoretical background}\label{s.tb}

In this section we will review in a systematic way 
the basic ideas used in the theoretical analysis of ultrafast carrier 
dynamics in semiconductors. The approach we are going to present is based on
the density-matrix formalism introduced in section \prr{ss.moc}.

\subsection{Physical system}\label{ss.ps}

In order to study the optical and transport properties of semiconductor 
bulk and hetero\-struc\-tu\-res, let us consider a gas of carriers in a 
crystal under the action of an applied electromagnetic field. The carriers 
will experience their mutual interaction as well as the interaction with the 
phonon modes of the crystal.
Such physical system can be described by the following Hamiltonian:
\begin{equation}\label{eq1}
{\bf H} = {\bf H}_c + {\bf H}_p + {\bf H}_{cc} + {\bf H}_{cp} 
+ {\bf H}_{pp}\ .
\end{equation}
The first term describes the noninteracting-carrier system 
in the presence of the external electromagnetic field while the 
second one refers to the free-phonon system. The last three terms describe 
many-body contributions: they refer, respectively, to carrier-carrier, 
carrier-phonon, and phonon-phonon interactions.

In order to discuss their explicit form, let us introduce the usual 
second-quantization field operators
${\bf \Psi}^\dagger({\bf r})$ and ${\bf \Psi}({\bf r})$. They describe, 
respectively, the creation and the destruction of a carrier in ${\bf r}$.
In terms of the above field operators the carrier Hamiltonian ${\bf H}_c$ 
can be written as
\begin{equation}\label{eq2}
{\bf H}_c = \int d{\bf r} {\bf \Psi}^\dagger({\bf r})
\left[
{\left(-i\hbar\nabla_{\bf r} -{e\over c} {\bf A}({\bf r},t)\right)^2
\over 2 m_\circ} + e\varphi({\bf r},t) + V^l({\bf r})
\right] 
{\bf \Psi}({\bf r})\ . 
\end{equation}
Here, $V^l({\bf r})$ denotes the periodic potential due to the perfect crystal 
while ${\bf A}({\bf r},t)$ and $\varphi({\bf r},t)$ denote, respectively, 
the vector and scalar potentials corresponding to the external 
electromagnetic field. Since we are interested in the electrooptical 
properties as well as in the ultrafast dynamics of photoexcited carriers, 
the electromagnetic field acting on the crystal ---and the corresponding 
electromagnetic potentials--- will be the sum 
of two different contributions: the high-frequency laser field responsible 
for the ultrafast optical excitation and the additional electromagnetic 
field acting on the photoexcited carriers on a longer time-scale.
More specifically, by denoting with the labels $1$ and $2$ these two 
contributions, we can write
\begin{equation}\label{eq3}
{\bf A}({\bf r},t) = {\bf A}_1({\bf r},t) + {\bf A}_2({\bf r},t)\ , \qquad
\varphi({\bf r},t) = \varphi_1({\bf r},t) + \varphi_2({\bf r},t)
\end{equation}
and recalling that
\begin{equation}\label{eq4}
{\bf E}({\bf r},t) = -{1\over c} {\partial\over\partial t} {\bf A}({\bf r},t)
-\nabla_{\bf r} \varphi({\bf r},t)\ , \qquad 
{\bf B}({\bf r},t) = \nabla_{\bf r}\times{\bf A}({\bf r},t)
\end{equation}
we have
\begin{equation}\label{eq5}
{\bf E}({\bf r},t) = {\bf E}_1({\bf r},t) + {\bf E}_2({\bf r},t)\ , \qquad
{\bf B}({\bf r},t) = {\bf B}_1({\bf r},t) + {\bf B}_2({\bf r},t)\ .
\end{equation}
Equation (\prr{eq4}), which gives the electromagnetic fields in terms of 
the corresponding vector and scalar potentials, 
reflects the well known gauge freedom: there is an infinite number of
possible combinations of ${\bf A}$ and $\varphi$ which give rise to the same 
electromagnetic field $\{{\bf E}, {\bf B}\}$.
We will use such freedom of choice for the laser field (term $1$): we 
assume a homogeneous (space-independent) laser field ${\bf E}_1(t)$ fully 
described by the scalar potential
\begin{equation}\label{eq6}
\varphi_1({\bf r},t) = -{\bf E}_1(t) \cdot {\bf r}\ .
\end{equation}
This assumption, which corresponds to the well known dipole approximation, is 
well justified as long as the space-scale of interest is small compared to 
the light wavelength.
The explicit form of the laser field considered here is 
\begin{equation}\label{eq7}
E_1(t) = E^+(t) + E^-(t) =
E^{ }_\circ(t) e^{i\omega_L t} +
E^*_\circ(t) e^{-i\omega_L t}\ ,
\end{equation}
where $E_\circ(t)$ is the amplitude of the light field and
$\omega_L$ denotes its central frequency.

With this particular choice of the electromagnetic potentials describing 
the laser field, the Hamiltonian in (\prr{eq2}) can be rewritten as
\begin{equation}\label{eq8}
{\bf H}^{ }_c = {\bf H}^\circ_c + {\bf H}^{ }_{cl}\ ,
\end{equation}
where 
\begin{equation}\label{eq9}
{\bf H}^\circ_c = \int d{\bf r} {\bf \Psi}^\dagger({\bf r})
\left[
{\left(-i\hbar\nabla_{\bf r} -{e\over c} {\bf A}_2({\bf r},t)\right)^2
\over 2 m_\circ} + e\varphi_2({\bf r},t) + V^l({\bf r})
\right] 
{\bf \Psi}({\bf r})
\end{equation}
describes the carrier system in the crystal under the action of the 
electromagnetic field $2$ only, while
\begin{equation}\label{eq10}
{\bf H}^{ }_{cl} = e \int d{\bf r} {\bf \Psi}^\dagger({\bf r})
\varphi_1({\bf r},t) 
{\bf \Psi}({\bf r})
\end{equation}
describes the carrier-light (cl) interaction due to the laser photoexcitation.

In analogy with the carrier system, by denoting with 
$b^\dagger_{{\bf q},\lambda}$ and $b^{ }_{{\bf q},\lambda}$ the creation 
and destruction operators for a phonon of mode $\lambda$ and wavevector 
${\bf q}$, the free-phonon Hamiltonian takes the form
\begin{equation}\label{eq11}
{\bf H}_p = \sum_{{\bf q}\lambda} 
\hbar\omega_{{\bf q}\lambda}
b^\dagger_{{\bf q}\lambda} b^{ }_{{\bf q}\lambda}\ ,
\end{equation}
where $\omega_{{\bf q}\lambda}$ is the dispersion relation for the phonon 
mode $\lambda$.

Let us now discuss the explicit form of the many-body contributions.
The carrier-carrier interaction is described by the two-body 
Hamiltonian
\begin{equation}\label{eq12}
{\bf H}_{cc} = {1\over 2} \int d{\bf r} \int d{\bf r}' 
{\bf \Psi}^\dagger({\bf r}) {\bf \Psi}^\dagger({\bf r}') 
V_{cc}({\bf r}-{\bf r}') 
{\bf \Psi}({\bf r}') {\bf \Psi}({\bf r}) \ ,
\end{equation}
where $V_{cc}$ denotes the Coulomb potential.

Let us now introduce the carrier-phonon interaction Hamiltonian
\begin{equation}\label{eq13}
{\bf H}_{cp} = \int d{\bf r} 
{\bf \Psi}^\dagger({\bf r}) V_{cp}({\bf r}) {\bf \Psi}({\bf r})\ ,
\end{equation}
where
\begin{equation}\label{eq14}
V_{cp} = \sum_{{\bf q}\lambda} 
\left[
\tilde{g}^{ }_{{\bf q}\lambda} b^{ }_{{\bf q}\lambda} 
e^{i{\bf q}\cdot{\bf r}}
+ 
\tilde{g}^*_{{\bf q}\lambda} b^\dagger_{{\bf q}\lambda} 
e^{-i{\bf q}\cdot{\bf r}}
\right]
\end{equation}
is the electrostatic phonon potential induced by the lattice vibrations.
Here, the explicit form of the coupling function 
$\tilde{g}_{{\bf q}\lambda}$ 
depends on the particular phonon mode $\lambda$ (acoustic, optical, etc.)
as well as on the coupling mechanism considered (deformation potential, 
polar coupling, etc.).

Let us finally discuss the phonon-phonon contribution ${\bf H}_{pp}$. The 
free-phonon Hamiltonian ${\bf H}_p$ in (\prr{eq11}), 
which describes a system of noninteracting phonons, by definition 
accounts only for the harmonic part of the lattice potential. 
However, non-harmonic contributions of the 
interatomic potential can play an important role in determining the lattice
dynamics in highly excited systems \prcite{Kash89}, since they are 
responsible for the decay of optical phonons into phonons of lower frequency.
In our second-quantization picture, these non-harmonic contributions 
can be described in terms of a phonon-phonon interaction which induces, 
in general, transitions between free-phonon states.
Here, we will not discuss the explicit form of the phonon-phonon 
Hamiltonian ${\bf H}_{pp}$ responsible for such a decay. 
We will simply assume that such phonon-phonon interaction is so efficient to 
maintain the phonon system in thermal equilibrium. This corresponds to 
neglecting hot-phonon effects \prcite{Poetz83}.

It is well known that the coordinate representation used so far is not the 
most convenient one in describing the electron dynamics within a periodic 
crystal. In general, it is more convenient to employ the 
representation given by the eigenstates of the noninteracting-carrier 
Hamiltonian ---or a part of it--- since it automatically accounts for 
some of the symmetries of the system.
For the moment we will simply consider an orthonormal basis set 
$\{\phi_n({\bf r})\}$ without specifying which part of the 
Hamiltonian is diagonal in such representation. This will allow us to 
write down equations valid in any quantum-mechanical representation.
Since the noninteracting-carrier Hamiltonian is, in general, a function of
time, also the basis functions $\phi_n$ may be time-dependent.
Here, the label $n$ denotes, in general, a set of discrete and/or continuous 
quantum numbers.
In the absence of electromagnetic field, the above wavefunctions will 
correspond to the well known Bloch states of the crystal and the index $n$ 
will reduce to the wavevector ${\bf k}$ plus the band (or subband) 
index $\nu$. 
In the presence of a homogeneous magnetic
field the eigenfunctions $\phi_n$ may instead correspond to Landau states. 
Finally, for the case of a constant and homogeneous electric field, 
there exist two equivalent representations: the accelerated Bloch states 
and the Wannier-Stark picture. Such equivalence results to be of crucial 
importance in understanding the relationship between Bloch oscillations and
Wannier-Stark localization and, for this reason, it will be discussed in
more detail in section \prr{ss.qmp}.

Let us now reconsider the system Hamiltonian introduced so far in terms of 
such $\phi_n$ representation. As a starting point, we may expand the 
second-quantization field operators in terms of the new wavefunctions:
\begin{equation}\label{eq15}
{\bf \Psi}({\bf r}) = \sum_n \phi^{ }_n({\bf r}) a^{ }_n\ , \qquad 
{\bf \Psi}^\dagger({\bf r}) = \sum_n \phi^*_n({\bf r}) a^\dagger_n\ .
\end{equation}
The above expansion defines the new set of second-quantization operators 
$a^\dagger_n$ and $a^{ }_n$; They describe, respectively, the creation and 
destruction of a carrier in state $n$.

For the case of a semiconductor structure (the only one 
considered here), the energy spectrum of the 
noninteracting-carrier Hamiltonian (\prr{eq9}) ---or a part of it---
is always characterized by two well-separated energy regions: the valence 
and the conduction band. 
Also in the presence of an applied electromagnetic field, the periodic 
lattice potential $V^l$ gives rise to a large energy gap. Therefore, we 
deal with two energetically well-separated regions, which suggests the 
introduction of the usual electron-hole picture. 
This corresponds to a separation of the set of states $\{\phi_n\}$ 
into conduction states $\{\phi^e_i\}$ 
and valence states $\{\phi^h_j\}$.
Thus, also the creation (destruction) operators 
$a^\dagger_n$ ($a^{ }_n$) introduced in equation (\prr{eq15}) will be 
divided into creation (destruction) electron and hole operators: 
$c^\dagger_i$ ($c^{ }_i$) and $d^\dagger_j$ ($d^{ }_j$).
In terms of the new electron-hole picture, the expansion 
(\prr{eq15}) is given by:
\begin{eqnarray}\label{eq16}
{\bf \Psi}({\bf r}) = && \sum_i \phi^e_i({\bf r}) c^{ }_i 
+ \sum_j \phi^{h *}_j({\bf r}) d^\dagger_j \nonumber \\[1ex]
{\bf \Psi}^\dagger({\bf r}) = && \sum_i \phi^{e *}_i({\bf r}) c^\dagger_i 
+ \sum_j \phi^h_j({\bf r}) d^{ }_j \ .
\end{eqnarray}
If we now insert the above expansion into equation (\prr{eq9}), the 
noninteracting-carrier Hamiltonian takes the form
\begin{equation}\label{eq17}
{\bf H}^\circ_c = \sum_{ii'} \epsilon^e_{ii'} c^\dagger_i c^{ }_{i'} + 
\sum_{jj'} \epsilon^h_{jj'} d^\dagger_j d^{ }_{j'}
= {\bf H}^\circ_e + {\bf H}^\circ_h\ ,
\end{equation}
where 
\begin{equation}\label{eq18}
\epsilon^{e/h}_{ll'} = \pm
\int d{\bf r} \phi^{e/h *}_l({\bf r}) \left[
{\left(-i\hbar\nabla_{\bf r} -{e\over c} {\bf A}_2\right)^2
\over 2 m_\circ} + e\varphi_2 + V^l 
- \epsilon_\circ
\right] \phi^{e/h}_{l'}({\bf r})
\end{equation}
are just the matrix elements of the Hamiltonian in the 
$\phi$-representation. The  $\pm$ sign refers, respectively, to electrons 
and holes while $\epsilon_\circ$ denotes the conduction-band 
edge.
Here, we neglect any valence-to-conduction band coupling due to the 
external electromagnetic field and vice versa. 
This is well fulfilled for the systems and field-regimes  we are going to 
discuss in this paper.
As already pointed out, the above Hamiltonian may be time-dependent. 
We will discuss this aspect in the following section, where 
we will derive our set of kinetic equations.

Let us now write in terms of our electron-hole representation the 
carrier-light interaction Hamiltonian (\prr{eq10}):
\begin{equation}\label{eq19}
{\bf H}_{cl} = -\sum\limits_{i,j} \left[
\mu^{eh}_{ij} E^-(t) c^{\dagger}_i d^{\dagger}_j
+\mu^{eh *}_{ij} E^+(t) d^{ }_j c^{ }_i
\right]\ .
\end{equation}
The above expression has been obtained within the well known rotating-wave 
approximation \prcite{Haug93} 
by neglecting intraband transitions, absent for the case of 
optical excitations.
Here, $\mu^{eh}_{ij}$ denotes the optical 
dipole matrix element between states $\phi^e_i$ and $\phi^h_j$.

Similarly, the carrier-carrier Hamiltonian (\prr{eq12}) can be rewritten as:
\begin{eqnarray}\label{eq20}
{\bf H}_{cc} = && \frac{1}{2}\sum\limits_{i_1i_2i_3i_4} V^{cc}_{i_1i_2i_3i_4}
c^{\dagger}_{i_1}c^{\dagger}_{i_2}c_{i_3}c_{i_4}\nonumber\\
&& + \frac{1}{2}\sum\limits_{j_1j_2j_3j_4} V^{cc}_{j_1j_2j_3j_4}
d^{\dagger}_{j_1}d^{\dagger}_{j_2}d_{j_3}d_{j_4}\nonumber\\
&& - \sum\limits_{i_1i_2j_1j_2} V^{cc}_{i_1j_1j_2i_2}
    c^{\dagger}_{i_1}d^{\dagger}_{j_1}d_{j_2}c_{i_2},
\end{eqnarray}
where
\begin{equation}\label{eq21}
V^{cc}_{l_1l_2l_3l_4} = \int d{\bf r} \int d{\bf r}'
\phi^*_{l_1}({\bf r})\phi^*_{l_2}({\bf r}')
V^{cc}({\bf r}-{\bf r}') 
\phi_{l_3}({\bf r}')\phi_{l_4}({\bf r})
\end{equation}
are the Coulomb matrix elements within our $\phi$-representation.
The first two terms describe the repulsive electron-electron and hole-hole 
interaction while the last one describes the attractive electron-hole 
interaction.
Here, we neglect terms that do not conserve the number of electron-hole 
pairs, i.e. impact-ionization and Auger-recombination 
processes \prcite{Quade94}, 
as well as the interband exchange interaction. 
This monopole-monopole approximation is justified as long as the
exciton binding energy is small compared to the energy gap.

Finally, let us rewrite the carrier-phonon interaction Hamiltonian 
introduced in equation (\prr{eq13}):
\begin{eqnarray}\label{eq22}
{\bf H}_{cp} = && \sum\limits_{ii',{\bf q}\lambda} 
\left[ g^{e}_{ii',{\bf q}\lambda} 
c_i^\dagger b^{ }_{{\bf q}\lambda} c^{ }_{i'} +
g^{e *}_{ii',{\bf q}\lambda} 
c^\dagger_{i'} b^\dagger_{{\bf q}\lambda} c^{ }_i \right] \nonumber \\
&& - \sum\limits_{jj',{\bf q}\lambda} \left[
g^h_{jj',{\bf q}\lambda} 
d^\dagger_j b^{ }_{{\bf q}\lambda} d^{ }_{j'} +
g^{h *}_{jj',{\bf q}\lambda} 
d^\dagger_{j'} b^\dagger_{{\bf q}\lambda} d^{ }_j \right]
\end{eqnarray}
with
\begin{equation}\label{eq23}
g^{e/h}_{ll',{\bf q}\lambda} = 
\tilde{g}^{ }_{{\bf q}\lambda} \int d{\bf r} 
\phi^{e/h *}_l({\bf r})
e^{i{\bf q}\cdot{\bf r}}
\phi^{e/h}_{l'}({\bf r})\ .
\end{equation}
In equation (\prr{eq22}) we can clearly recognize four different 
contributions corresponding to electron and hole phonon absorption and 
emission.
\subsection{Kinetic description}\label{ss.kd}

Our kinetic description of the ultrafast carrier dynamics in 
semiconductors is based on the density-matrix 
formalism introduced in section \prr{ss.moc}. 
Since this approach has been reviewed and discussed in several papers 
\prcite{Kuhn97,Rossi92b} and 
text-books \prcite{Shah96,Haug93},
here we will simply recall in our notation the kinetic equations
relevant for the analysis of carrier dynamics in semiconductor 
heterostructures, generalizing 
the approach presented in \prcite{Kuhn97} to the case of a time-dependent 
quantum-mechanical representation.

The set of kinetic variables is the same considered in 
\prcite{Kuhn97}. Given our 
electron-hole representation $\{\phi^e_i\}, \{\phi^h_j\}$, 
we will consider the intraband electron and hole single-particle density 
matrices
\begin{equation}\label{eq24}
 f^e_{ii'} = \left\langle c^\dagger_i c^{ }_{i'} \right\rangle\ , \qquad
f^h_{jj'} = \left\langle d^\dagger_j d^{ }_{j'} \right\rangle
\end{equation}
as well as the corresponding interband density matrix
\begin{equation}\label{eq25}
 p^{ }_{ji} = \left\langle d^{ }_j c^{ }_i \right\rangle\ .
\end{equation}
Here, the diagonal elements $f^e_{ii}$ and $f^h_{jj}$ correspond to 
the electron and hole distribution functions of the Boltzmann theory  
while the non-diagonal terms describe intraband polarizations. 
On the contrary, the interband density-matrix elements $p^{ }_{ji}$ describe 
interband (or optical) polarizations. 

In order to derive the set of kinetic equations, i.e. the equations of 
motion for the above kinetic variables, the standard procedure starts by 
deriving the equations of motion for the electron and hole 
operators introduced in (\prr{eq16}):
\begin{equation}\label{eq26}
c^{ }_i = \int d{\bf r} \phi^{e *}_i({\bf r}) {\bf \Psi}({\bf r})\ , \qquad
d^{ }_j = \int d{\bf r} \phi^{h *}_j({\bf r}) {\bf \Psi}^\dagger({\bf r})\ . 
\end{equation}
By applying the Heisenberg equation of motion for the field operator 
${\bf \Psi}$, i.e.
\begin{equation}\label{eq27}
\frac{d}{dt} {\bf \Psi} = \frac{1}{i\hbar} 
\left[{\bf \Psi},{\bf H}\right]\ ,
\end{equation}
it is easy to obtain the following equations of motion:
\begin{eqnarray}\label{eq28}
\frac{d}{dt} c^{ }_i = && \frac{1}{i\hbar} \left[c^{ }_i,{\bf H}\right] + 
\frac{1}{i\hbar}\sum_{i'} Z^e_{ii'} c^{ }_{i'} \nonumber \\[1ex]
\frac{d}{dt} d^{ }_j = && \frac{1}{i\hbar} \left[d^{ }_j,{\bf H}\right] + 
\frac{1}{i\hbar}\sum_{j'} Z^h_{jj'} d^{ }_{j'}\
\end{eqnarray}
with
\begin{equation}\label{eq29}
Z^{e/h}_{ll'} = i\hbar \int d{\bf r} \left(\frac{d}{dt}
\phi^{e/h *}_l({\bf r})\right) \phi^{e/h}_{l'}({\bf r})\ .
\end{equation}
As for the case of equation (\prr{eq17}), here we neglect again 
valence-to-conduction band coupling and vice versa. 
Compared to the more conventional Heisenberg equations of motion, 
they contain an extra-term, the last one. It accounts for the 
possible time dependence of our $\phi$-re\-pre\-sen\-ta\-tion which will 
induce 
transitions between different states according to the matrix elements 
$Z^{ }_{ll'}$.

By combining the above equations of motion with the definitions of the kinetic 
variables in (\prr{eq24}-\prr{eq25}), 
the resulting set of kinetic 
equations can be schematically written as:
\begin{eqnarray}\label{eq30}
\frac{d}{dt} f^e_{i_1i_2} = && \frac{d}{dt} f^e_{i_1i_2}\Biggl|_{\bf H} + 
\frac{d}{dt} f^e_{i_1i_2}\Biggl|_\phi \nonumber \\[1ex]
\frac{d}{dt} f^h_{j_1j_2} = && \frac{d}{dt} f^h_{j_1j_2}\Biggl|_{\bf H} + 
\frac{d}{dt} f^h_{j_1j_2}\Biggl|_\phi \nonumber \\[1ex]
\frac{d}{dt} p^{ }_{j_1i_1} = && \frac{d}{dt} p^{ }_{j_1i_1}\Biggl|_{\bf H} + 
\frac{d}{dt} p^{ }_{j_1i_1}\Biggl|_\phi \ .
\end{eqnarray}
They exhibit the same structure of the equations of motion (\prr{eq28}) 
for the 
electron and hole creation and destruction operators: a first term induced 
by the system Hamiltonian ${\bf H}$ (which does not account for the time 
variation of the basis states) and a second one induced by the 
time dependence of the 
basis functions $\phi$.

Let us start discussing this second term, whose explicit form is:
\begin{eqnarray}\label{eq31}
\frac{d}{dt} f^e_{i_1i_2}\Biggl|_\phi = && \frac{1}{i\hbar}
\sum\limits_{i_3i_4}\left[Z^e_{i_2i_4}\delta_{i_1i_3} -
Z^e_{i_3i_1}\delta_{i_2i_4}\right] f^e_{i_3i_4} 
\nonumber \\[1ex]
\frac{d}{dt} f^h_{j_1j_2}\Biggl|_\phi = && \frac{1}{i\hbar}
\sum\limits_{j_3j_4}\left[Z^h_{j_2j_4}\delta_{j_1j_3} -
Z^h_{j_3j_1}\delta_{j_2j_4} \right] f^h_{j_3j_4} \nonumber \\[1ex]
\frac{d}{dt} p^{ }_{j_1i_1}\Biggl|_\phi = && \frac{1}{i\hbar} 
\sum\limits_{i_2j_2}\left[Z^h_{j_1j_2}\delta_{i_1i_2} +
Z^e_{i_1i_2}\delta_{j_1j_2} \right] p^{ }_{j_2i_2} \ .
\end{eqnarray}
As we will see in section \prr{ss.qmp}, these contributions 
play a central role for the description of Zener tunneling within the 
vector-potential representation.

Let us now come to the first term. This, in turn, is the sum of different 
contributions, corresponding to the various parts of the Hamiltonian.
The total Hamiltonian can be regarded as the sum of two 
terms, a single-particle contribution plus a many-body one:
\begin{equation}\label{eq32}
{\bf H} = {\bf H}_{sp} + {\bf H}_{mb} = 
\left({\bf H}^\circ_c + {\bf H}_{cl} + {\bf H}_p\right) + 
\left({\bf H}_{cc} + {\bf H}_{cp}
+ {\bf H}_{pp}\right)\ .
\end{equation}
The explicit form of the time evolution due to the single-particle 
Hamiltonian ${\bf H}_{sp}$ (non-interacting carriers plus carrier-light 
interaction plus free phonons) is given by:
\begin{eqnarray}\label{eq33}
\frac{d}{dt} f^e_{i_1i_2}\Biggl|_{sp} = && \frac{1}{i\hbar} 
\bigg\{\sum\limits_{i_3i_4}\left[\epsilon^e_{i_2i_4}\delta_{i_1i_3} -
\epsilon^e_{i_3i_1}\delta_{i_2i_4}\right] f^e_{i_3i_4} \nonumber \\
&& + \sum\limits_{j_1}\left[U_{i_2j_1} p^*_{j_1i_1}
-U^*_{i_1j_1} p^{ }_{j_1i_2}\right] \bigg\} \nonumber \\
\frac{d}{dt} f^h_{j_1j_2}\Biggl|_{sp} = && \frac{1}{i\hbar} \bigg\{
\sum\limits_{j_3j_4}\left[\epsilon^h_{j_2j_4}\delta_{j_1j_3} -
\epsilon^h_{j_3j_1}\delta_{j_2j_4} \right] f^h_{j_3j_4} \nonumber \\
&& + \sum\limits_{i_1}\left[U_{i_1j_2} p^*_{j_1i_1}
-U^*_{i_1j_1} p^{ }_{j_2i_1} \right] \bigg\} \nonumber \\ 
\frac{d}{dt} p^{ }_{j_1i_1}\Biggl|_{sp} = && \frac{1}{i\hbar} \bigg\{
\sum\limits_{i_2j_2}\left[\epsilon^h_{j_1j_2}\delta_{i_1i_2} +
\epsilon^e_{i_1i_2}\delta_{j_1j_2} \right] p^{ }_{j_2i_2} \nonumber \\
&& + \sum\limits_{i_2j_2} U^{ }_{i_2j_2}
\left[\delta_{i_1i_2}\delta_{j_1j_2} - f^e_{i_2i_1}\delta_{j_1j_2}
- f^h_{j_2j_1}\delta_{i_1i_2} \right] \bigg\}
\end{eqnarray}
with $U_{i_1j_1} = -\mu^{eh}_{i_1j_1} E^-(t)$.

This is a closed set of equations, 
which is a consequence of the single-particle nature 
of ${\bf H}_{sp}$. 
In addition, we stress that the structure of the two contributions 
entering equation (\prr{eq30}) is very similar: 
one can include the contribution (\prr{eq31}) into (\prr{eq33}) by 
replacing $\epsilon$ with $\epsilon + Z$. 

Let us finally discuss the contributions due to the many-body part of the 
Hamiltonian: carrier-carrier and carrier-phonon interactions (the 
phonon-phonon one is not explicitly considered here).
As discussed in \prcite{Kuhn97}, 
for both interaction mechanisms one can 
derive a hierarchy of equations involving higher-order density matrices. 
In order to close such equations with respect to our set of kinetic 
variables, approximations are needed.
The lowest-order contributions to our equations of motion are given by 
first-order terms in the many-body Hamiltonian: Hartree-Fock level. Since 
we will neglect coherent-phonon states, the only Hartree-Fock 
contributions will come from carrier-carrier interaction. 
They simply result in a renormalization 
\begin{equation}\label{eq34}
\Delta\epsilon^{e/h}_{l_1l_2} =
- \sum\limits_{l_3l_4} V^{cc}_{l_1l_3l_2l_4} f^{e/h}_{l_3l_4}
\end{equation}
of the single-particle energy matrices $\epsilon^{e/h}$ 
as well as in a renormalization
\begin{equation}\label{eq36}\
\Delta U_{i_1j_1} = - \sum\limits_{i_2j_2} 
V^{cc}_{i_1j_1j_2i_2} p^{ }_{j_2i_2}
\end{equation}
of the external field $U$. 
\footnote{The explicit form of the renormalization terms considered in this 
paper accounts for the Fock contributions only, i.e. no Hartree terms. The
general structure of Hartree-Fock contributions, relevant for the case of a
strongly non-homogeneous system, is discussed in \prcite{Kuhn97}.}
We stress that the Hartree-Fock approximation, which consists in 
factorizing average values of four-point operators into products of 
two density matrices, is independent from the quantum-mechanical 
picture.
It is then clear that the above kinetic equations are valid in
any quantum-mechanical representation.

All the contributions to the system dynamics discussed so far describe a 
fully coherent dynamics, i.e. no scattering processes.
In order to treat incoherent phenomena, e.g. energy relaxation and 
dephasing, one has to go one step further in the perturbation expansion 
taking into account also second-order contributions (in the perturbation 
Hamiltonian ${\bf H}_{mb}$).
The derivation of these higher-order contributions, discussed in 
\prcite{Kuhn97},
will not be repeated here. Again, 
in order to obtain a closed set of equations (with 
respect to our set of kinetic variables (\prr{eq24}-\prr{eq25})) additional 
approximations are needed, namely
the mean-field and Markov approximations.
As for the Hartree-Fock case, 
the mean-field approximation allows to write the various higher-order 
density matrices as products of single-particle ones. 
The Markov approximation allows to eliminate the additional higher-order 
kinetic variables, e.g. phonon-assisted density matrices, providing a 
closed set of equations still local in time, i.e. no memory effects 
\prcite{Brunetti89,TranThoai93,Sayed94,Schilp94,Meden95}.
This last approximation is not performed in the 
quantum-kinetic theory discussed in \prcite{Kuhn97}
where, in addition to our single-particle 
variables, one considers two-particle 
\prcite{Quade94}
and phonon-assisted 
\prcite{Schilp94}
density matrices. 

While the mean-field approximation is representation-independent, this is 
unfortunately not the case for the Markov limit.
This clearly implies that the validity of the Markov approximation is 
strictly related to the quantum-mechanical representation considered.

The above kinetic description, based on intra- and interband density 
matrices, allows us to evaluate any single-particle quantity.
In particular, for the analysis of the ultrafast carrier dynamics in 
photoexcited semiconductors two physical quantities play a central role: 
the intra- and interband 
total (or macroscopic) polarizations:
\begin{equation}\label{eq37}
P^{e/h}(t) = \sum_{ll'} M^{e/h}_{ll'} f^{e/h}_{l'l}(t)\ , \qquad
P^{eh} = \sum_{ij} \mu^{eh}_{ij} p^{ }_{ji}(t)\ ,
\end{equation}
where $M^{e/h}$ and $\mu^{eh}$ denote, respectively, the intra- and interband 
dipole matrix elements in our 
$\phi$-re\-pre\-sen\-ta\-tion.
The time derivative of the intraband polarization $P^{e/h}$ 
describes the radiation field induced by the Bloch-oscillation dynamics 
(which for a superlattice structure is in the TeraHertz range) 
while the Fourier transform of the interband (or optical) polarization 
$P^{eh}$ 
provides the linear and non-linear optical response of the system.

\section{Coherent ultrafast dynamics in bulk semiconductors}\label{s.bulk}

In this section, we will discuss the dominant role played by 
optically-induced phase coherence on the ultrafast generation and 
relaxation of photoexcited carriers in bulk semiconductors. 
In particular, we will compare the 
description based on the theoretical approach discussed above with the 
more conventional picture based on the Boltzmann theory. 

\subsection{Bloch model}\label{ss.sbe}

In order to study the ultrafast carrier dynamics
in bulk systems, let us consider a two-band semiconductor model.
For the case of a homogeneous system, the only relevant terms of 
the single-particle density matrix in ${\bf k}$-space are the diagonal 
ones.
This property, due to the translational symmetry of the 
Hamiltonian, reduces the set of kinetic variables (\prr{eq24}-\prr{eq25}) 
to the following electron and hole distribution functions 
(intraband density-matrix elements)
\begin{equation}\label{eq1bulk}
f^e_{{\bf k}} = \left\langle c^\dagger_{{\bf k}} 
c^{ }_{{\bf k}} \right\rangle\ , \qquad
f^h_{{\bf -k}} = \left\langle d^\dagger_{{\bf -k}} 
d^{ }_{{\bf -k}} \right\rangle
\end{equation}
together with the corresponding polarizations (interband density-matrix 
elements) \footnote{Here, the standard 
electron-hole picture introduced in section \prr{ss.ps} has 
been applied to our plane-wave states.
In particular, due to 
the charge-conjugation symmetry, the hole states are still labeled in terms
of the corresponding valence-electron states, i.e. ${\bf k}^h \equiv 
-{\bf k}^e$.}
\begin{equation}\label{eq2bulk}
p^{ }_{{\bf k}} = \left\langle d^{ }_{{\bf -k}} 
c^{ }_{{\bf k}} \right\rangle\ .
\end{equation}
The explicit form of the kinetic equations (\prr{eq30}) within the above 
two-band picture is given by
\begin{eqnarray}\label{eq3bulk}
\frac{d}{dt} f_{\bf k}^e = && g_{\bf k}(t) - \sum_{\bf k'} \left[ W^e_{\bf k'k}
f_{\bf k}^e \left( 1 - f_{\bf k'}^e \right) - W^e_{\bf kk'} f_{\bf k'}^e
\left( 1 - f_{\bf k}^e \right) \right] \nonumber \\
\frac{d}{dt} f_{\bf k}^h = && g_{\bf -k}(t) 
- \sum_{\bf k'} \left[ W^h_{\bf k'k}
f_{\bf k}^h \left( 1 - f_{\bf k'}^h \right) - W^h_{\bf kk'} f_{\bf k'}^h
\left( 1 - f_{\bf k}^h \right) \right] \nonumber \\
\frac{d}{dt} p_{\bf k}^{ } = && \frac{1}{i\hbar}\left[ \left(
\epsilon_{\bf k}^e + \epsilon_{\bf -k}^h \right) p_{\bf k}^{ } +
U_{\bf k}\left( 1 - f_{\bf k}^e - f_{\bf
-k}^h \right) \right] \nonumber \\
&& {} 
- \sum_{\bf k'} \left[ W^p_{\bf k'k}
p_{\bf k}^{ } - W^p_{\bf kk'} p_{\bf k'}^{ } \right] \ ,
\end{eqnarray}
with the generation rate
\begin{equation}\label{eq4bulk}
g_{\bf k} = \frac{1}{i\hbar} \left[ U^{ }_{\bf k}p^*_{\bf k} 
- U^*_{\bf k} p^{ }_{\bf k}\right] \ .
\end{equation}
Here, the incoherent contributions are treated within the usual Markov 
limit as described in \prcite{Kuhn97,Kuhn92b,Haas96}. 
Within such approximation scheme, the incoherent contributions to the 
polarization dynamics exhibit the same structure as for the distribution 
functions in terms of the following in- and out-scattering rates 
\begin{equation}\label{eq5bulk}
W^p_{\bf k'k} = \frac{1}{2} \sum_{\nu=e,h} \left[
 W^\nu_{\bf k'k}  \left( 1 - f_{\bf k'}^\nu \right) + W^\nu_{\bf k'k}
f_{\bf k'}^\nu \right] \ ,
\end{equation}
where $W^{e/h}_{\bf k'k}$ are the usual scattering rates of the 
semiclassical Boltzmann theory. 
This Boltzmann-like structure of the scattering term in the polarization 
equation is the starting point of the generalized Monte Carlo method 
discussed in section \prr{s.int} 
\prcite{Kuhn92a,Kuhn92b,Rossi93,Rossi94,Haas96}.

The above kinetic equations are known as {\it semiconductor Bloch equations} 
(SBE)\footnote{In the absence of Coulomb correlation and scattering, the 
SBE in (\prr{eq3bulk}) reduce to the optical Bloch equations introduced in 
section \prr{ss.moc}, i.e. they describe a collection of independent 
(non-interacting) two-level systems.}.
As we can see, here the carrier photogeneration is a two-step process: 
The external field $U$ induces a coherent polarization $p$ which, in 
turn, generates electron-hole pairs via its coupling with the field 
according to equation (\prr{eq4bulk}).
The generation rate is thus determined by the interband polarization 
$p_{\bf k}$ which, in turn, is influenced by the various density-dependent 
scattering mechanisms, e.g. carrier-carrier processes. 
Therefore, in contrast to the semiclassical case discussed below, 
the generation rate within the SBE model is clearly density dependent. 

\subsection{Boltzmann model}\label{ss.be}

The conventional Boltzmann model is obtained from the above SBE by 
performing an adiabatic elimination of the interband polarization 
$p_{\bf k}$, i.e. by inserting a formal solution of the polarization 
equation into the generation rate (\prr{eq4bulk}) and then performing a 
Markov limit with respect to the electron and hole distribution functions 
\prcite{Kuhn92b,Haas96}.
This corresponds to assuming that the carrier distributions are slowly-varying 
functions on the time-scale of the laser photoexcitation, i.e. the 
carrier generation and relaxation are treated as independent processes.
Within such approximation scheme, 
except for phase-space-filling effects, the generation rate entering the 
semiclassical Boltzmann equations (BE) for electrons and holes 
is fully determined by the temporal and
spectral characteristics of the laser pulse.
Contrary to the SBE case, 
all effects related to the optically induced phase 
coherence ---and to its temporal decay---
are neglected. 

\subsection{Coherent carrier photogeneration}\label{ss.cpg}

In order to compare the different models of carrier dynamics in photoexcited 
semiconductors, let us review some simulated experiments \prcite{Kuhn95} 
based on the generalized Monte Carlo solution of the SBE 
\prcite{Kuhn92a,Kuhn92b,Rossi93,Rossi94,Haas96}
discussed in section \prr{s.int}.
The corresponding BE has been solved, for comparison, by a standard 
ensemble Monte Carlo (EMC) simulation \prcite{Jacoboni89,Rossi92a}. 
Both BE and SBE have been solved numerically for the case of GaAs bulk
excited by a 150~fs laser pulse. 

In figure \prr{fig1bulk} the generation rate at different times 
as obtained from the SBE
is plotted as a function of the carrier wave vector for three different 
densities. 
At the lowest density the behavior is 
essentially the same as in the case without carrier-carrier
scattering:
Energy-time uncertainty leads to an initially very broad 
generation rate; with increasing time the line narrows and, in the tails, 
exhibits negative parts due to a stimulated recombination of carriers
initially generated off-resonance \prcite{Kuhn92b,Haas96}. 
After the pulse, the 
distribution function of the generated carriers is in good agreement with 
the BE result. Scattering processes destroy the 
coherence between electrons and holes which is necessary for the
stimulated recombination processes. As a consequence, with increasing 
density the negative tails are strongly reduced and the generation remains 
broad for all times resulting in a much broader carrier distribution than 
in the BE case.

\subsection{Comparison with experiments}\label{cwe}

Hot carrier luminescence has proven to be a powerful technique to study the
ultrafast dynamics of photoexcited carriers in 
semiconductors \prcite{Elsaesser91}. 
In band-to-acceptor (BA) luminescence experiments 
\prcite{Ulbrich89,Snoke92,Kane94,Kash95}, due to their high 
sensitivity, carrier densities as low as 
several $10^{13}$~cm$^{-3}$ have been reached. Thus, the transition between
a dynamics dominated by carrier-phonon scattering to a dynamics dominated 
by carrier-carrier scattering (with its consequences for the generation 
process discussed above) is experimentally 
accessible. 

In the BA luminescence experiment reviewed here 
\prcite{Leitenstorfer94b,Kuhn95,Leitenstorfer96b},
a 3~$\mu$m thick GaAs layer doped with Be acceptors of a 
concentration of $3 \times 10^{16}$ cm$^{-3}$
is excited at a photon energy of 1.73~eV by transform limited 150 fs pulses
from a mode-locked Ti:sapphire laser. Time-integrated luminescence 
spectra are recorded \prcite{Leitenstorfer94b}. 
The right column of figure \prr{fig2bulk} shows measured BA luminescence 
spectra for three different values of the carrier density. At low density 
we observe an initial peak of the generated carriers (marked bold) and
pronounced replicas due to the emission of an integer number of optical 
phonons. With increasing density the peaks become broader and at the 
highest density only a slight structure related to the phonons is 
still visible.

These measured spectra are compared 
with corresponding simulated experiments 
based on the BE and SBE models \prcite{Kuhn95} 
(left and 
middle column in figure \prr{fig2bulk}). 
We find several pronounced differences between the two models. 
In the semiclassical model the 
unrelaxed peak is clearly visible up to the highest density in contrast to 
the coherent model, where, in agreement with the experiment, 
this peak is strongly broadened. At
the higher densities the semiclassical spectra exhibit an increase in the 
broadening of subsequent replicas which is not present in the 
Bloch-model calculations as well as in the experimental results.

To illustrate this difference quantitatively, in figure \prr{fig3bulk} 
we have plotted the full 
width at half maximum (FWHM) of the three highest peaks in the spectra. 
From the semiclassical model we obtain
a strong increase in the density dependence of subsequent peaks. 
This behavior can be 
easily understood: Due to the emission time of an optical phonon of about 
$150$~fs, the carriers populate subsequent replicas at increasing times and, 
therefore, the efficiency of carrier-carrier scattering processes in 
broadening the peaks increases. However, the measured spectra exhibit 
a different behavior which is quantitatively reproduced by the SBE model: 
Already the unrelaxed peak exhibits a strong increase with 
increasing density and the density dependence of all replicas is 
approximately the same. The reason is that the broadening of the
generation process as discussed above obviously dominates over the 
broadening due to subsequent scattering processes. As discussed in 
\prcite{Leitenstorfer96b}, this is due to the fact that the 
generation process is strongly influenced by the decay of the interband 
polarization which, in turn, is 
due to electron plus hole scattering, while 
the broadening of the electron distribution during the relaxation process 
is due to electron scattering only. \footnote{
For the case of bulk GaAs considered here, due to the different electron 
and heavy-hole effective masses, the hole-hole scattering is about 5 times 
larger than the electron-electron one.}

The above theoretical and experimental analysis 
constitutes a clear demonstration of the 
importance of dephasing processes for the analysis of luminescence spectra.
The density dependence of the spectra in the interesting transition region 
between carrier-phonon and carrier-carrier dominated dynamics 
can only be explained by a coherent modeling of 
the carrier generation including the dynamics of the interband 
polarization. The main mechanism determining the width of the peaks is the 
broadening of the generation rate.
It can be shown that for band-to-acceptor spectra the 
broadening of the recombination process, neglected in the present model, 
plays a minor role \prcite{Kuhn95}.
However, for band-to-band spectra also this phenomenon should be taken into 
account. 
For the case discussed above the background due to band-to-band luminescence 
has been found to be of negligible importance \prcite{Leitenstorfer96b}.

\section{Bloch oscillations and Wannier-Stark localization in superlattices}
\label{s.sl}

Ever since the initial applications of quantum mechanics to the dynamics of
electrons in solids, the analysis of Bloch electrons moving in a 
homogeneous electric field has been of central importance.
By employing semiclassical arguments, in 1928 Bloch \prcite{Bloch28} 
demonstrated that, a wave packet given by a superposition of single-band states 
peaked about some quasimomentum, $\hbar{\bf k}$, moves 
with a group velocity given by the gradient of the energy-band function 
with respect to the quasimomentum and that the rate of change of the 
quasimomentum is proportional to the applied field ${\bf F}$. 
This is often referred to as the ``acceleration theorem'':
\begin{equation}\label{eq1sl}
\hbar\dot{\bf k} = e{\bf F}\ .
\end{equation}
Thus, in the absence of interband tunneling and scattering processes, 
the quasimomentum of a Bloch 
electron in a homogeneous and static electric field will be uniformly 
accelerated into the next Brillouin zone in a repeated-zone scheme (or 
equivalently undergoes an Umklapp process back in to the first zone). 
The corresponding motion of the Bloch electron through the periodic 
energy-band structure, shown in figure \prr{fig1sl}, is called ``Bloch 
oscillation''; It is characterized 
by an oscillation period $\tau_B = h/eFd$, where $d$ denotes the lattice 
periodicity in the field direction. 

There are two mechanisms impeding a fully periodic motion: interband 
tunneling and scattering processes.
Interband tunneling is an intricate problem and still at the center of a 
continuing debate.
Early calculations of the tunneling probability into other bands in which 
the electric field is represented by a time-independent scalar potential 
were made by Zener \prcite{Zener34} using a Wentzel-Kramers-Brillouin 
generalization of Bloch functions, 
by Houston \prcite{Houston40} using accelerated Bloch states (Houston 
states),
and subsequently by Kane \prcite{Kane59} and Argyres \prcite{Argyres62} 
who employed the crystal-momentum representation.
Their calculations lead to the conclusion that the tunneling 
rate per Bloch period is much less than unity for electric fields up to 
$10^6$V/cm for 
typical band parameters corresponding to elemental or compound 
semiconductors.

Despite the apparent agreement among these calculations, the 
validity of employing the crystal-momentum representation or Houston 
functions to describe electrons moving in a 
non-periodic (crystal plus external field) potential has been 
disputed. 
The starting point of the controversy was the original paper by 
Wannier \prcite{Wannier60}. 
He pointed out that, 
due to the translational symmetry of the crystal potential, 
if $\phi({\bf r})$ is an 
eigenfunction of the scalar-potential Hamiltonian (corresponding to the 
perfect crystal plus the external field)
with eigenvalue 
$\epsilon$, 
then any $\phi({\bf r}+n{\bf d})$ is also an
eigenfunction with eigenvalue 
$\epsilon+n \Delta\epsilon$, where $\Delta\epsilon = e F d$ is the 
so-called Wannier-Stark splitting
(${\bf d}$ being the primitive 
lattice vector along the field direction). 
He concluded that the translational symmetry of the crystal gives rise
to a discrete energy spectrum, the so-called Wannier-Stark ladder. The 
states corresponding to these equidistantly spaced levels are localized 
states, as schematically shown in figure \prr{fig2sl} for the case of a 
semiconductor superlattice.

The existence of such energy quantization was disputed 
by Zak \prcite{Zak68a}, who pointed out that 
for the case of an infinite crystal the scalar potential 
$-{\bf F}\cdot{\bf r}$ is not bounded, which implies a continuous energy 
spectrum.
Thus, the main point of the controversy was related to the 
existence (or absence) of Wannier-Stark ladders. 
More precisely, the point 
was to decide if interband tunneling (neglected in the original calculation
by Wannier \prcite{Wannier60})
is so strong to destroy the 
Wannier-Stark energy quantization (and the corresponding Bloch oscillations) 
or not.

It is only during the last decade that this controversy came to an end.
From a theoretical point of view, most of the formal problems 
related to the non-periodic nature of the scalar potential (superimposed to 
the periodic crystal potential) were finally removed by using 
a vector-potential representation of the applied 
field \prcite{Kittel63,Krieger86}.
Within such vector-potential picture, 
upper boundaries for the interband 
tunneling probability have been established at a rigorous level, which show
that an electron may execute a number of Bloch oscillations before 
tunneling out of the band \prcite{Krieger86,Nenciu91}, in 
qualitatively good agreement with the earlier predictions of Zener and Kane 
\prcite{Zener34,Kane59}.

The second mechanism impeding a fully periodic motion is scattering by 
phonons, impurities, etc. (see figure \prr{fig1sl}). 
This results in lifetimes shorter than the Bloch 
period $\tau_B$ for all reasonable values of the electric 
field, so that Bloch oscillations should not be observable in conventional 
solids. 

In superlattices, however, the situation is much more favourable because of 
the smaller Bloch period $\tau_B$ resulting from the small width of the 
mini-Brillouin zone in the field direction \prcite{Bastard89}.

Indeed, the existence of Wannier-Stark ladders as well as 
Bloch oscillations in 
superlattices has been confirmed by a number of recent 
experiments \prcite{Shah96}.
The photoluminescence and photocurrent measurements of the biased 
GaAs/GaAlAs superlattices performed by Mendez and coworkers 
\prcite{Mendez88},
together with the electroluminescence experiments by Voisin and 
coworkers \prcite{Voisin88},
provided the earliest 
evidence of the field-induced Wannier-Stark ladders in superlattices. 
A few years later, Feldmann and coworkers \prcite{Feldmann92} were able to 
measure Bloch oscillations in the time domain through a 
four-wave-mixing experiment originally suggested by 
von Plessen and Thomas \prcite{vonPlessen92}.
A detailed analysis of the Bloch oscillations in the four-wave-mixing signal 
(which reflects the interband dynamics) has been also performed by Leo and 
coworkers \prcite{Leo92,Leisching94}.

In addition to the above interband-polarization analysis, Bloch 
oscillations have been also detected by monitoring the intraband 
polarization which, in turn, is reflected by anisotropic changes in the 
refractive index \prcite{Shah96}. 
Measurements based on transmittive electrooptic sampling (TEOS) have been 
performed by Dekorsy and coworkers \prcite{Dekorsy94,Dekorsy95}.  
Finally Bloch oscillations were recently measured through a direct 
detection of the TeraHertz (THz) radiation in 
semiconductor superlattices \prcite{Waschke93,Roskos94}.

\subsection{Two equivalent pictures}\label{ss.qmp}

Let us now apply the theoretical approach presented in section \prr{s.tb} 
to the case of a semiconductor superlattice in the presence of an 
uniform (space-independent) electric field.
The non-interacting carriers within the superlattice crystal will then be
described by the Hamiltonian ${\bf H}^\circ_c$ in equation (\prr{eq9}),
where now the electrodynamic potentials ${\bf A}_2$ and $\varphi_2$ 
(in the following simply denoted with 
${\bf A}$ and $\varphi$) correspond 
to a homogeneous electric field 
${\bf E}_2({\bf r},t) = {\bf F}(t)$.

As pointed out in section \prr{ss.ps}, the natural quantum-mechanical 
representation is given by the eigenstates of this Hamiltonian:
\begin{equation}\label{eq2sl}
\left[
{\left(-i\hbar\nabla_{\bf r} -{e\over c} {\bf A}({\bf r},t)\right)^2
\over 2 m_\circ} + e\varphi({\bf r},t) + V^l({\bf r})
\right] \phi_n({\bf r}) = \epsilon_n \phi_n({\bf r})\ .
\end{equation}
However, due to the gauge freedom discussed in section \prr{ss.ps},
there is an infinite number of
possible combinations of ${\bf A}$ and $\varphi$ 
---and therefore of possible Hamiltonians---
which describe the same 
homogeneous electric field ${\bf F}(t)$.
In particular, one can identify two independent choices:
the vector-potential gauge
\begin{equation}\label{eq3sl}
{\bf A}({\bf r},t) = -c \int_{t_\circ}^t {\bf F}(t') dt'\ , \qquad 
\varphi({\bf r},t) = 0
\end{equation}
and the scalar-potential gauge 
\begin{equation}\label{eq4sl}
{\bf A}({\bf r},t) = 0\ , \qquad 
\varphi({\bf r},t) = -{\bf F}(t) \cdot {\bf r}
\end{equation}

As shown in \prcite{Rossi97}, the two independent choices correspond, respectively, 
to the well known Bloch-oscillation and Wannier-Stark pictures. 
They simply reflect two equivalent quantum-mechanical representations and, 
therefore, any physical phenomenon can be described in both
pictures. 

More specifically, within the vector-potential picture (\prr{eq3sl}), 
the eigenfunctions $\phi_n$ in (\prr{eq2sl}) are the so-called
accelerated Bloch states (or Houston states) 
\prcite{Houston40,Kittel63,Krieger86}. 
As discussed in \prcite{Rossi97}, such time dependent representation 
constitutes the natural basis for the description of Bloch 
oscillations, i.e. it provides a rigorous quantum-mechanical derivation 
of the acceleration theorem (\prr{eq1sl}), thus showing that this is not a
simple semiclassical result. \footnote{The acceleration theorem 
(\prr{eq1sl}) 
and the corresponding Bloch-oscillation dynamics are usually regarded as a 
semiclassical result compared to the Wannier-Stark picture. On the 
contrary, they correspond to two different fully quantum-mechanical 
pictures.}
Within such representation, Bloch oscillations are fully described by 
the diagonal terms of the intraband density matrix (\prr{eq24}) 
(semiclassical distribution functions). Therefore, non-diagonal elements, 
describing phase-coherence between different Bloch states, do not 
contribute to the intraminiband dynamics. However, they are of crucial 
importance for the description of interminiband dynamics, i.e. 
field-induced Zener tunneling, which in this Bloch-state representation 
originates from the time variation of our basis states (see equation 
(\prr{eq31})).

On the contrary, within the scalar-potential picture (\prr{eq4sl}), 
the eigenfunctions $\phi_n$ in (\prr{eq2sl}) are the well-known 
Wannier-Stark states \prcite{Wannier60}. Contrary to the previous Bloch 
picture, within such representation the intraminiband Bloch dynamics originates 
from a quantum interference between different Wannier-Stark states, thus 
involving non-diagonal elements of the intraband density matrix 
(\prr{eq24}).
\subsection{Some simulated experiments}\label{ss.sse}

In this section, we will review recent simulated experiments of the 
ultrafast carrier dynamics in semiconductor 
superlattices \prcite{Rossi95a,Rossi95b,Meier95b,Je95,Koch95,Rossi96a}.
They are based on a generalized Monte Carlo 
solution \prcite{Kuhn92a,Kuhn92b,Rossi93,Rossi94,Haas96} 
of the set of kinetic 
equations (so-called semiconductor Bloch equations) derived in 
section \prr{ss.kd}. 
In this case, the Bloch representation discussed in section \prr{ss.qmp} has 
been employed limiting the set of interband density-matrix elements in 
(\prr{eq25}) to the diagonal ones, i.e. $i = j$.
In addition, incoherent scattering processes have been treated within the 
usual Markov limit as discussed in \prcite{Kuhn97,Kuhn92b,Haas96}.
Within such approximation scheme, the explicit form of the kinetic 
equations (\prr{eq30}) coincides with the SBE (\prr{eq3}) 
(obtained for the bulk case) \prcite{Rossi97}, 
provided we replace the wavevector ${\bf k}$ with ${\bf k}\nu$, 
$\nu$ being the superlattice miniband index.

In the simulated experiments reviewed here the following superlattice model 
has been employed:
The energy dispersion and the corresponding 
wavefunctions along the growth direction ($k_\parallel$)
are computed within the well known
Kronig-Penney model \prcite{Bastard89}, while 
for the in-plane direction ($k_\perp$) 
an effective-mass model has been used.
Starting from these three-dimensional wavefunctions 
$\phi^\circ_{{\bf k}\nu}$, the 
various carrier-carrier as well as 
carrier-phonon matrix elements are numerically computed (see 
equations (\prr{eq21}) and (\prr{eq23})).
They are, in general, functions
of the various miniband indices and depend separately 
on $k_\parallel$ and $k_\perp$, 
thus fully reflecting the anisotropic nature  of the 
superlattice structure.

Only coupling to GaAs bulk phonons has been considered.
This, of course, is a simplifying approximation which neglects any
superlattice effect on the phonon dispersion, such as
confinement of optical modes in the wells and in the barriers,
and the presence of interface modes \prcite{Ruecker92,Molinari94}.
However, while these modifications
have important consequences for phonon spectroscopies
(like Raman scattering), they are far less decisive for transport
phenomena. 
\footnote{Indeed, by now it is well known \prcite{Molinari94}
that the total scattering rates are sufficiently well reproduced
if the phonon spectrum is assumed to be bulk-like.}

We will start discussing the scattering-induced damping of 
Bloch oscillations. In particular, we will show that in the low-density limit 
this damping is mainly determined by optical-phonon 
scattering \prcite{Rossi95a,Rossi95b} 
while at high densities the main mechanism
responsible for the suppression of Bloch 
oscillations is found to be carrier-carrier scattering \prcite{Rossi96a}.

This Bloch-oscillation analysis in the time domain is also confirmed by its
counterpart in the frequency domain. As pointed out in 
section \prr{ss.qmp}, 
the presence of Bloch oscillations, due to a negligible scattering 
dynamics, should correspond to Wannier-Stark energy quantization.
This is confirmed by the simulated optical-absorption spectra, 
which clearly show the presence of the field-induced Wannier-Stark 
ladders\prcite{Koch95}.
\subsubsection{Bloch-oscillation analysis}\label{sss.boa}

All the simulated experiments presented in this section refer to 
the superlattice structure considered in Ref.~\prcite{Meier95b}:
$111$ \AA\ GaAs wells and $17$ \AA\ Al$_{0.3}$Ga$_{0.7}$As barriers. 
For such a structure
there has been experimental evidence for a THz-emission 
from Bloch oscillations \prcite{Roskos94}.

In the first set of simulated experiments an initial distribution of 
photoexcited carriers (electron-hole pairs) is generated by a
$100$ fs Gaussian laser pulse in resonance with the first-miniband 
exciton ($\hbar \omega_L \approx 1540$ meV).
The strength of the applied electric field is assumed to be $4$ kV/cm, which 
corresponds to a Bloch period $\tau_B = h/eFd$ of about $800$ fs.

In the low-density limit (corresponding to a weak laser excitation),
incoherent scattering processes do not alter the Bloch-oscillation dynamics. 
This is due to the following reasons:
In agreement with recent experimental \prcite{Roskos94,vonPlessen94} and 
theoretical \prcite{Rossi95a,Rossi95b,Meier95b} investigations,
for superlattices characterized by a miniband width
smaller than the LO-phonon energy 
---as for the structure considered here---
and for laser 
excitations close to
the band gap, at low temperature carrier-phonon scattering is not 
permitted. 
Moreover, 
in this low-density regime carrier-carrier scattering plays no role: Due 
to the quasi-elastic nature of Coulomb collisions, 
in the low-density limit the majority of the 
scattering processes 
is characterized by a very small
momentum transfer; As a consequence, the momentum relaxation along the 
growth direction is negligible.
As a result, on this picosecond time-scale the carrier system exhibits a 
coherent Bloch-oscillation dynamics, i.e. a negligible scattering-induced 
dephasing.
This can be clearly seen from the time evolution of the carrier distribution 
as a function of $k_\parallel$
(i.e. averaged over $k_{\perp}$) 
shown in figure \prr{fig3sl}. 
During the laser photoexcitation ($t = 0$) 
the carriers are generated
around $k_\parallel = 0$, where the transitions are close to resonance with
the laser excitation. According to the
acceleration theorem (\prr{eq1sl}), the electrons are then shifted in
${\bf k}$-space. When the carriers reach the border of the first
Brillouin zone they are Bragg reflected.
After about $800$ fs, corresponding to the Bloch period $\tau_B$, 
the carriers have completed one oscillation
in ${\bf k}$-space. 
As expected, the carriers execute Bloch oscillations without
loosing the synchronism of their motion by scattering.
This is again shown in figure \prr{fig3sl}, where we have plotted: (b) the mean 
kinetic energy, (c) the current, and (d) its time derivative
which is proportional to the emitted far field, i.e. the THz-radiation.
(It can be shown that, by neglecting Zener tunneling,  the intraband 
polarization $P^{e/h}$ in equation (\prr{eq37}) is proportional to the 
current.)
All these three quantities exhibit oscillations characterized by the same 
Bloch period $\tau_B$.
Due to the finite width of the carrier distribution in ${\bf k}$-space 
(see figure \prr{fig3sl}(a)), 
the amplitude of the oscillations of the kinetic energy is somewhat smaller 
than the miniband width. 
Since for this excitation condition the 
scattering-induced dephasing is negligible,
the oscillations of the current are symmetric around zero, which implies 
that the time average of the current is equal to zero, i.e. no dissipation.

As already pointed out, this ideal Bloch-oscillation regime is typical of a
laser excitation close to gap in the low-density limit. Let us now discuss, 
still at low densities, the case of a laser photoexcitation high in the 
band.
Figure \prr{fig4sl}(a) shows the THz-signal as obtained from a set of simulated 
experiments corresponding to different laser excitations \prcite{Meier95b}.
The different traces correspond to the emitted
THz-signal for increasing excitation energies. We clearly notice the 
presence of Bloch oscillations in all cases. However, the oscillation 
amplitude and decay (effective damping) is excitation-dependent.

For the case of a laser excitation resonant with the first-miniband exciton 
considered above (see figure \prr{fig3sl}),
we have a strong THz-signal. 
The amplitude of the signal decreases when the excitation energy is
increased. Additionally,
there are also some small changes in the phase of the oscillations,
which are induced by the electron-LO phonon scattering.

When the laser energy comes into resonance with the transitions between the
second electron and hole minibands ($\hbar \omega_L \approx 1625$ meV),
the amplitude of the THz-signal increases again. The corresponding 
THz-transients show an initial part, which is strongly damped
and some oscillations for longer times that are much less damped.
For a better understanding of these results,
we show in figure \prr{fig4sl}(b) the individual
THz-signals, originating from
the two electron and two heavy-hole minibands for the excitation with
$\hbar \omega = 1640$ meV. 
The Bloch oscillations performed by the electrons within the second miniband
are strongly damped due to intra- and interminiband LO-phonon scattering
processes \prcite{Rossi95a,Meier95b}.
Since the width of this second miniband ($45$ meV) is somewhat larger than the
LO-phonon energy, also intraminiband scattering is possible, whenever the
electrons are accelerated into the high-energy region
of the miniband.  
The THz-signal originating from electrons within the first miniband
shows an oscillatory behavior with a small amplitude and a phase 
which is determined by the time the electrons need to relax
down to the bottom of the band.

At the same time, 
the holes in both minibands exhibit undamped Bloch oscillations,
since the minibands are so close in 
energy that for these excitation conditions no LO-phonon emission
can occur. The analysis shows that
at early times the THz-signal is mainly determined by
the electrons within the
second miniband. At later times the observed signal is
due to heavy holes and electrons within the first miniband.

The above theoretical analysis closely resembles 
experimental observations obtained for a superlattice structure 
very similar to the one modelled here \prcite{Roskos94}. 
In these experiments, evidence for THz-emission 
from Bloch oscillations has been reported. For some 
excitation conditions these oscillations are associated
with resonant excitation of the second miniband. 
The general behavior of the
magnitude of the signals, the oscillations and the damping
are close to the results shown in figure \prr{fig4sl}.

Finally, in order to study the density dependence of the Bloch-oscillation 
damping, let us go back to the case of laser excitations close to gap.
Figure \prr{fig5sl}(a) shows the total (electrons plus holes) THz-radiation 
as a function of time for three different carrier densities. 
With increasing carrier density, carrier-carrier scattering
becomes more and more important: Due to Coulomb screening, the 
momentum transfer in a carrier-carrier scattering increases (its typical 
value being comparable with the screening wavevector). This can be seen 
in figure \prr{fig5sl}(a), where for increasing carrier densities we realize an 
increasing damping of the THz-signal. 
However, also for the highest carrier density considered here we deal with 
a damping time of the order of $700$ fs, which is much larger than the 
typical dephasing time, i.e. the decay time of the interband polarization,
 associated with carrier-carrier scattering.
The dephasing time is typically 
 investigated by means of four-wave-mixing (FWM)
measurements and such multi-pulse experiments can be simulated as 
well \prcite{Lohner93,Leitenstorfer94a}.
From a theoretical point of view, a qualitative estimate of the dephasing 
time is given by the decay time of the ``incoherently summed'' polarization 
(ISP) \prcite{Kuhn92b}.
Figure \prr{fig5sl}(b) 
shows such ISP as a function of time for the same three carrier 
densities of figure \prr{fig5sl}(a).
As expected, the decay times are always much smaller than the 
corresponding damping times of the THz-signals (note the 
different time-scale in figures \prr{fig5sl}(a) and (b)).
This difference, discussed in more detail in \prcite{Rossi96a,Rossi97}, 
can be understood as follows:
The fast decay times of figure \prr{fig5sl}(b) reflect the 
interband dephasing, i.e. the
sum of the electron and hole scattering rates. In particular, for the 
Coulomb interaction this means the sum of electron-electron, electron-hole,
and hole-hole scattering.
As for the case of bulk GaAs discussed in section \prr{s.bulk}, 
this last contribution is known to dominate and determines the dephasing 
time-scale.
On the other hand, the total THz-radiation in figure \prr{fig5sl}(a) 
is the sum of the 
electron and hole contributions. However, due to the small value of the 
hole miniband width compared to the electron one, the electron 
contribution will dominate.
This means that the THz damping in figure \prr{fig5sl}(a) mainly reflects 
the damping of the electron contribution. 
This decay, in turn, reflects the intraband dephasing of electrons which is
due to electron-electron and electron-hole scattering only, i.e. no hole-hole 
contributions.

From the above analysis we can conclude that the decay time 
of the THz-radiation due to carrier-carrier scattering differs considerably
from the corresponding dephasing times obtained from a FWM experiment:
The first one is a measurement of the intraband dephasing while the second 
one reflects the interband dephasing.
\footnote{We stress that this difference between intraband and interband 
dephasing in superlattices is the same discussed in section \prr{s.bulk}
for the case of bulk semiconductors, 
where the broadening of the photoexcited carrier distribution is mainly 
determined by the decay of the interband polarization (interband dephasing) 
while the subsequent energy broadening of the electron distribution
is due to electron scattering only (intraband dephasing)
(see figure \prr{fig2bulk}).}
\subsubsection{Optical-absorption analysis}\label{sss.oaa}

Let us now discuss the frequency-domain counterpart of the 
Bloch-oscilla\-tion picture considered so far.
Similar to what happens in the time domain, for sufficiently high 
electric fields, i.e. when the
Bloch period $\tau_B = h/eFd$ becomes smaller than
the dephasing time, 
the optical spectra of the superlattice are expected to exhibit the 
frequency-domain counterpart of the
Bloch oscillations, i.e. the Wannier-Stark energy quantization discussed in
section \prr{ss.qmp}.
In the absence of Coulomb interaction, the Wannier-Stark ladder absorption 
increases as a function of the photon energy in a step-like fashion.
These steps are equidistantly spaced. Their spacing, named Wannier-Stark 
splitting, is proportional to the applied electric field.

The simulated linear-absorption spectra corresponding to a superlattice 
structure with 
$95$ \AA\ GaAs wells and $15$ \AA\ Al$_{0.3}$Ga$_{0.7}$As barriers 
are shown in figure \prr{fig6sl} \prcite{Koch95}.
As we can see, 
the Coulomb interaction gives rise to excitonic peaks in the absorption 
spectra and introduces couplings between these Wannier-Stark states.
Such exciton peaks, which are no longer equidistantly spaced, 
are often referred to as  
excitonic Wannier-Stark ladders \prcite{Dignam90} of the 
superlattice. 

Since for the superlattice structure considered in this simulated 
experiment \prcite{Rossi95a,Koch95} the combined 
miniband width is larger than the typical 
two- and three-dimensional exciton binding energies,
it is possible to investigate the quasi-three 
dimensional absorption behavior
of the delocalized miniband states as well as localization
effects induced by the electric field.

For the free-field case, the electron and hole states are completely
delocalized in our three-dimensional ${\bf k}$-space.
The perturbation induced by the application of a low field 
(here $\approx 5$ kV/cm), couples the
states along the field direction and in the spectra
the Franz-Keldysh effect,
well known from bulk materials \prcite{Haug93}, appears:
one clearly notices oscillations which increase in amplitude with the field
and shift with $F^{2/3}$ from the $n = 0$ and $n = 1$ levels
toward the center of the combined miniband. 

For increasing field the potential drop over the distance of a
few quantum wells eventually exceeds the miniband width and the 
electronic states 
become more and more localized. Despite the field-induced
energy difference $n eFd$,
the superlattice potential is equal for quantum wells separated by $nd$. 
Therefore,
the spectra decouple into a series of peaks
corresponding to the excitonic ground states of the individual
electron-hole Wannier-Stark levels.
Each Wannier-Stark transition contributes to the absorption
with a pronounced $1-s$ exciton peak, plus
higher bound exciton and continuum states. The oscillator
strength of a transition $n$ is proportional
to the overlap between electron and hole wavefunctions centered
at quantum wells $n'$ and $n+n'$, respectively. 
The analysis shows that this oscillator strength
is almost exclusively determined by the amplitude of the electron
wavefunction in quantum well $n'+n$ since for fields in the
Wannier-Stark regime
the hole wavefunctions are almost completely localized over one
quantum well due to their high effective mass
(see figure \prr{fig2sl}). Thus, the
oscillator strengths of transitions
to higher $|n|$ become smaller with increasing $|n|$ and field.    

At high fields (here $>\approx 8$ kV/cm) the separation between the peaks 
is almost equal to $neFd$.
For example, the peak of the $n=0$ transition
which is shifted by the 
Wannier-Stark exciton binding energy with respect 
to the center of the combined miniband, demonstrates that the
increasing localization also increases the 
exciton binding energy.  This increased 
excitonic binding reflects the gradual transition from
a three- to a two-dimensional behavior, discussed in detail in 
\prcite{Meier95b}. 

For intermediate fields there is an interplay between the
Wannier-Stark and the Franz-Keldysh effect.
Coming from high fields, first the Wannier-Stark peaks are
modulated by the Franz-Keldysh oscillations.
However, as soon as the separation $eFd$ between neighboring peaks becomes 
smaller than their spectral widths,
the peaks can no longer be resolved individually so that only
the Franz-Keldysh structure remains.   
\section{Coulomb-correlation effects in semiconductor quantum wires}
\label{s.qwr}

The importance of Coulomb-correlation effects in optical spectra of
semiconductors and their dependence on dimensionality has now been long
recognized \prcite{review_excitons}. 
More recently, increasing interest has been devoted to one-dimensional (1D) 
systems \prcite{review_wires}, prompted by 
promising advances in quantum-wire fabrication 
and application, e.g. quantum-wire lasers.
The main goal is to achieve structures with improved optical 
efficiency as compared to two-dimensional (2D) and three-dimensional (3D) ones. 
A common argument in favour of this effort is based on the  
well known van Hove divergence in the 1D joint density-of-states (DOS),  
which is expected to give rise to very sharp peaks in the optical spectra 
of 1D structures. Such prediction is however based on free-carrier 
properties of ideal 1D systems and it ignores Coulomb-correlation 
effects. 

Early theoretical investigations by Ogawa and coworkers 
\prcite{Ogawa91a,Ogawa91b}, 
based on a 1D single-subband model of the wires,
showed that the inverse-square-root singularity in the 1D DOS 
at the band edge is smoothed when excitonic effects are taken into account. 

In this section, we will review recent theoretical results 
\prcite{Rossi96b,Rossi96c,Rossi96d} based on a full 3D
description of realistic quantum-wire structures made available 
by the present technology, such as structures obtained by epitaxial
growth on non-planar substrates (V-shaped wires) \prcite{Rinaldi94}
or by cleaved-edge quantum well overgrowth (T-shaped wires) \prcite{Sakaki96}.
On the one hand, this approach allows an accurate determination of 1D-exciton 
binding energies; on the other hand, it allows us to 
investigate also the non linear (gain) regime. 
In both cases, one finds a strong suppression of the 1D band-edge 
singularity, in agreement with previous results based on simplified 1D 
models \prcite{Ogawa91a,Ogawa91b}.

This theoretical approach \prcite{Rossi96b,Rossi96c} is based on the general 
kinetic theory presented in section \prr{s.tb}, which allows a full 3D 
description of Coulomb interaction within a multiminiband scheme.
In particular, we focus on the quasi-equilibrium regime where 
the solution of the coupled kinetic equations (\prr{eq30}) 
simply reduces to the solution of the interband-polarization equation. 
This is performed by direct numerical evaluation of the 
polarization eigenvalues and eigenvectors \prcite{Rossi96c}, which 
fully determine the absorption spectrum as well as
the exciton wavefunctions. The main ingredients entering 
this calculation are the single-particle energies and 
wavefunctions, obtained numerically for the 2D confinement 
potential deduced, e.g., from TEM as in \prcite{Rinaldi94}. 

The above theoretical scheme has been applied to realistic V- and T-shaped 
wire structures. In particular, here we show results for the 
GaAs/AlGaAs V-wire structure of Ref.~\prcite{Rinaldi94}, whose cross-section 
is shown in figure \prr{fig1qwr}.
\subsection{Linear response: excitonic-absorption 
regime}\label{ss.ear}

Let us start considering the optical response of the system in the 
low-density limit.
In figure \prr{fig2qwr} we show the
linear-absorption spectra obtained when taking into account the lowest wire 
transition only.
Results of our Coulomb-correlated (CC) approach are 
compared to those of the free-carrier (FC) model . 
As we can see, 
electron-hole correlation introduces two important effects: 
First, the excitonic peak arises below the onset of the continuum, with 
a binding energy of about $12$~meV, in excellent agreement with 
recent experiments \prcite{Rinaldi94}.
Second, one finds a strong suppression of the 1D DOS singularity.
A detailed analysis of the physical origin of such suppression 
\prcite{Rossi96b,Rossi96c}
has shown that the quantity which is mainly modified by CC is the oscillator 
strength (OS). 
In figure \prr{fig3qwr}(a) the ratio between the CC and FC OS is plotted as a 
function of 
excess energy 
(solid line). 
This ratio is always less than one and goes to zero at the band edge, and
reflects a sort of hole in the electron-hole correlation function $g(z)$ 
as shown in figure \prr{fig3qwr}(b).
Such vanishing behaviour is found to dominate the 1D DOS singularity 
and, as a result, the absorption spectrum exhibits 
a regular behaviour at the band edge (solid line in figure \prr{fig2qwr}).

The above analysis shows that 
also for realistic quantum-wire structures electron-hole correlation leads 
to a strong suppression of the 1D band-edge singularity in the 
linear-absorption spectrum, contrary to the 2D and 3D cases.
\subsection{Non-linear response: gain regime}\label{ss.gr}

Most of the potential quantum-wire applications, i.e. 1D lasers and 
modulators, operate in strongly non-linear-response regimes
\prcite{review_wires}.
In such conditions, the above linear-response analysis has to be 
generalized taking into account additional factors as:
(i) screening effects,
(ii) band renormalization,
(iii) space-phase filling.

Figure \prr{fig4qwr} reports quantitative results for non-linear absorption 
spectra of realistic V-shaped wire structures at different 
carrier densities at room temperature.
As a reference we also show the results obtained by including 
the lowest subband only [figure \prr{fig4qwr}(a)].
In the low-density limit (case A: $n = 10^4$ cm$^{-1}$) we clearly recognize 
the exciton peak. With increasing carrier density,
the strength of the excitonic absorption decreases due to phase-space 
filling and screening of the attractive electron-hole interaction, and 
moreover the band renormalization leads to a red-shift of the continuum.
At a density of $4*10^6$ cm$^{-1}$ (case D) the spectrum already exhibits a 
negative region corresponding to stimulated emission, i.e. gain regime. 
As desired, the well pronounced gain spectrum extends over a limited energy 
region (smaller than the thermal energy); However, its shape 
differs considerably from the ideal FC one (curve marked 
with diamonds in the same figure). In particular, 
the band-edge singularity in the ideal FC gain spectrum 
is clearly smeared out by electron-hole correlation. 
The overall effect is a broader and less 
pronounced gain region.

Finally, figure \prr{fig4qwr}(b) shows the non-linear spectra corresponding 
to the realistic case of a 12-subband V-shaped wire.
In comparison with the single-subband case [figure \prr{fig4qwr}(a)], 
the multisubband nature is found to play an important role 
in modifying the typical shape of the gain spectra, which for both CC and FC 
models result to extend over a range much larger 
than that of the single-subband case for the present wire geometry.
In addition, the Coulomb-induced suppression of the single-subband 
singularities, here also due to intersubband-coupling effects, tends to 
reduce the residual structures in the gain profile.
Therefore, even in the ideal case of a quantum wire with negligible 
disorder and scattering-induced broadening, our analysis indicates
that, for the typical structure considered, 
the shape of the absorption spectra over the whole density range 
differs considerably from the sharp FC spectrum of figure \prr{fig2qwr}.

This tells us that, in order to obtain sharp gain profiles,
one of the basic steps in quantum-wire technology is to produce structures 
with increased subband splitting. 
However, the disorder-induced inhomogeneous broadening, not considered 
here, is known to increase significantly the spectral broadening 
and this effect is expected to increase with increasing subband splitting. 
Therefore, extremely-high-quality structures (e.g. single-monolayer 
control) seem to be the only possible candidates for successful quantum-wire 
applications.

\section{Summary and conclusions}\label{s.suco}

A review of coherent phenomena in optically-excited 
semiconductors has been presented. 
Our analysis has allowed the identification of two classes of phenomena: 
{\it optically-induced} and {\it Coulomb-induced} coherent phenomena.
We have shown that both classes can be described in terms of a 
unique theoretical framework based on the density-matrix formalism. 
Due to its generality, such quantum-kinetic approach allows a realistic 
description of coherent as well as incoherent, i.e. phase-breaking, 
processes, thus providing quantitative information on the coupled 
---coherent vs. incoherent--- carrier dynamics in photoexcited 
semiconductors.

The primary goal of the paper was to discuss the concept of 
quantum-mechanical phase 
coherence as well as its relevance and implications on semiconductor physics 
and technology.
In particular, we have analyzed the dominant role played by optically 
induced phase coherence on the process of ultrafast carrier photogeneration. 
We have then discussed typical field-induced 
coherent phenomena in semiconductor superlattices, e.g. Bloch oscillations 
and the corresponding THz radiation, as well as their dephasing dynamics.
Finally, we have analyzed the dominant role played by Coulomb 
correlation on the linear and non-linear optical spectra of realistic 
quantum-wire structures, namely the Coulomb-induced suppression of ideal 
1D-band-edge singularities.

Our analysis shows that the conventional separation between 
coherent and incoherent regimes is no longer valid for most of the recent 
ultrafast optical experiments in semiconductors. The reason is that on such 
extremely short time-scales the carrier dynamics is the result of a strong 
interplay between coherence and relaxation, thus preventing a time-scale 
separation between photogeneration and relaxation dynamics. 
Similar considerations apply to the various Coulomb-induced phenomena: 
The strong modifications induced by Coulomb correlation on the optical 
response of low-dimensional semiconductors play a significant role on 
the phase-breaking relaxation dynamics as well.
It is thus evident that any theoretical analysis of the optical response 
of semiconductor heterostructures must provide a proper description of both 
coherent and incoherent phenomena as well as of their mutual coupling  
on the same kinetic level.
\section*{Acknowledgments}

I am particularly grateful to Tilmann Kuhn for his essential contribution 
in understanding most of the ideas and concepts discussed in this 
topical review.
I wish to thank Stephan W. Koch, 
Torsten Meier, and Peter Thomas, as well as Elisa Molinari, 
for their relevant contributions to 
the research activity reviewed in the paper. 
I am also grateful to Roberto Cingolani, Thomas Elsaesser, Alfred 
Leitenstorfer, and Peter E. Selbmann for stimulating 
and fruitful discussions. 

This work was supported in part by the EC Commission through the Network 
``ULTRAFAST''.
\bigskip
\figmacro{
\label{fig1bulk}
Generation rate of electrons as a function of $k$ at different
times and densities.
After Ref.~\prcite{Kuhn95}.
}
\figmacro{
\label{fig2bulk}
Calculated and measured band-to-acceptor spectra for different carrier
densities. 
After Ref.~\prcite{Kuhn95}.
}
\figmacro{
\label{fig3bulk}
Full width at half maximum of the unrelaxed peak and the first and
second phonon replica as a function of carrier density obtained from
the calculated and measured spectra. The lines connecting the theoretical 
values are meant as guides to the eye.
After Ref.~\prcite{Kuhn95}.
}
\figmacro{
\label{fig1sl}
Schematic illustration of the field-induced coherent motion of 
an electronic wavepacket initially created at the bottom of a miniband. 
Here, the width of the miniband exceeds the LO-phonon energy $E_{LO}$, 
so that LO-phonon scattering is possible. After Ref.~\prcite{vonPlessen94}.
}
\figmacro{
\label{fig2sl}
Schematic representation of the transitions from the valence to the 
conduction band of a superlattice in the Wannier-Stark localization regime. 
After Ref.~\prcite{Waschke93}.
}
\figmacro{
\label{fig3sl}
Full Bloch-oscillation dynamics corresponding to a laser 
photoexcitation resonant with the first-miniband exciton. (a) Time 
evolution of the electron distribution as a function of $k_\parallel$.
(b) Average kinetic energy, (c) current, and (d) THz-signal corresponding 
to the Bloch oscillations in (a).
}
\figmacro{
\label{fig4sl}
(a) Total THz-signals for eight different spectral positions of 
the exciting laser pulse: $1540, 1560,\dots, 1680$ meV (from bottom to top).
(b) Individual THz-signal of the electrons and holes in the different bands
for a central spectral position of the laser pulse of 1640 meV.
After Ref.~\prcite{Koch95}.
}
\figmacro{
\label{fig5sl}
(a) Total THz-radiation as a function of time;
(b) Incoherently-summed polarization as a function of time. After 
Ref.~\prcite{Rossi96a}.
}
\figmacro{
\label{fig6sl}
Absorption spectra for various static applied electric fields for a
GaAs/Al$_{0.3}$Ga$_{0.7}$As superlattice (well (barrier) width $95$ ($15$) 
\AA). The vertical displacements between any two spectra is proportional 
to the difference of the corresponding fields.
The Wannier-Stark transitions are labeled by numbers, the lower (higher) 
edge of the combined miniband by $E_0$ ($E_1$).
After Ref.~\prcite{Koch95}.
}
\figmacro{
\label{fig1qwr}
Typical cross-section of V-grooved wires derived from TEM micrographs and 
used to define the confinement potential entering the 2D single-particle 
Schr\"odinger equation.
The wire width along $y$ at the apex of the V is about $10$~nm.
The dashed lines identify the region (about $70 \times 25$~nm), where 
single-particle and excitonic wavefunctions are localized.
After Ref.~\prcite{Rossi96c}.
}
\figmacro{
\label{fig2qwr}
Linear-absorption spectra of realistic V-shaped quantum-wire structures 
obtained by including the first electron and hole subbands only.
Solid line: Coulomb correlated (CC) result; Dashed line: free-carrier 
(FC) result. 
After Ref.~\prcite{Rossi96d}.
}
\figmacro{
\label{fig3qwr}
(a) OS ratio and DOS vs. excess energy; (b) electron-hole 
correlation function.
After Ref.~\prcite{Rossi96d}.
}
\figmacro{
\label{fig4qwr}
Non-linear absorption spectra: (a) single-subband case; (b) realistic 
12-subband case.
After Ref.~\prcite{Rossi96d}.
}

\begin{references}

\bibitem{Shah96}
J. Shah, {\it Ultrafast Spectroscopy of Semiconductors and Semiconductor 
Nanostructures} (Springer, Berlin, 1996).
\bibitem{Allen75}
L. Allen, J.H. Eberly, {\it Optical Resonance and Two-Level Atoms} 
(Interscience, New York 1975).
\bibitem{Sargent77}
M. Sargent III, M.O. Scully, W.E. Lamb Jr., {\it Laser Physics} 
(Addison-Wesley, New York 1977).
\bibitem{Shen84}
Y.R. Shen, {\it The Principles of Nonlinear Optics} (Wiley, New York 1984).
\bibitem{Levenson88}
M.D. Levenson, S.S. Kano, {\it Introduction to Nonlinear Laser 
Spectroscopy} (Academic, Boston 1988).
\bibitem{Meystre91}
P. Meystre, M. Sargent III, {\it Elements of Quantum Optics}, 2nd edn. 
(Springer, Berlin, Heidelberg 1991).
\bibitem{Mills91}
D.L. Mills, {\it Nonlinear Optics} (Springer, Berlin, Heidelberg 1991).
\bibitem{Haug93}
H. Haug and S.W. Koch, {\it Quantum Theory of the Optical and Electronic 
Properties of Semiconductors}, 3rd Edn. (World Scientific, Singapore 1994).
\bibitem{Demtroder96}
W. Demtroeder, {\it Laser Spectroscopy}, 2nd edn. (Springer, Berlin, 
Heidelberg 1996).
\bibitem{Henneberger93}
F. Henneberger, S. Schmitt-Rink, E.O. G\"obel (eds.), {\it Optics of 
Semiconductor Nanostructures} (Akademie Verlag, Berlin 1993).
\bibitem{Phillips94}
R.T. Phillips (ed.) {\it Coherent Optical Processes in Semiconductors}, 
NATO ASI Series B: Physics, Vol. 30 (Plenum, New York 1994).
\bibitem{Hahn50}
E.L. Hahn, Phys. Rev.{ \bf 80}, 580 (1950).
\bibitem{Kurnit64}
N.A. Kurnit, I.D. Abella, and S.R. Hartmann, Phys. Rev. Lett. {\bf 13}, 
567 (1964).
\bibitem{Abella66}
I.D. Abella, N.A. Kurnit, and S.R. Hartmann, Phys. Rev. {\bf
141}, 391 (1966).
\bibitem{Shah69}
J. Shah and R.C. C. Leite, Phys. Rev. Lett. {\bf 22}, 1304 (1969).
\bibitem{Shank79}
C.V. Shank, R.L. Fork, R.F. Leheny, and J. Shah, Phys. Rev. Lett. {\bf 42}, 
112 (1979).
\bibitem{Shah87}
J. Shah, B. Deveaud, T.C. Damen, W.T. Tsang, A.C. Gossard, and P. Lugli,
Phys. Rev. Lett. {\bf 59}, 2222 (1987).
\bibitem{Elsaesser91}
T. Elsaesser, J. Shah, L. Rota, and P. Lugli, 
Phys. Rev. Lett. {\bf 66}, 1757 (1991).
\bibitem{Ulbrich89}
R.G. Ulbrich, J.A. Kash, and J.C. Tsang, 
Phys. Rev. Lett. {\bf 62}, 949 (1989).
\bibitem{Peterson90}
C.L. Peterson and S.A. Lyon, 
Phys. Rev. Lett. {\bf 65}, 760 (1990).
\bibitem{Snoke92}
D.W. Snoke, W.W. R\"uhle, Y.-C. Lu, and E. Bauser, 
Phys. Rev. Lett. {\bf 68}, 990 (1992).
\bibitem{Oudar84}
J.L. Oudar, A. Migus, D. Hulin, G. Grillon, J. Etchepare, and A. Antonetti,
Phys. Rev. Lett. {\bf 53}, 384 (1984).
\bibitem{Foing92}
J.-P. Foing, D. Hulin, M. Joffre, M.K. Jackson, J.-L. Oudar, C. Tanguy, and
M. Combescot, 
Phys. Rev. Lett. {\bf 68}, 110 (1992).
\bibitem{Leitenstorfer96a}
A. Leitenstorfer, C. F\"urst, A. Laubereau, W. Kaiser, G. Tr\"ankle, and G.
Weimann, Phys. Rev. Lett. {\bf 76}, 1545 (1996).
\bibitem{Lutgen96}
S. Lutgen, R. Kaindl, M. Woerner, T. Elsaesser, A. Hase, H. K\"unzel,
M. Gulia, D. Meglio, and P. Lugli, Phys. Rev. Lett. {\bf 77}, 3657 (1996). 
\bibitem{Joschko97}
M. Joschko, M. Woerner, T. Elsaesser, E. Binder, T. Kuhn, R. Hey, H. Kostial,
and K. Ploog, Phys. Rev. Lett.{ \bf 78}, 737 (1997).
\bibitem{Jacoboni89}
C. Jacoboni and P. Lugli, {\it The Monte Carlo Method for Semiconductor
Device Simulations} (Springer, Wien 1989).
\bibitem{Goodnick92}
S.M. Goodnick and P. Lugli, in {\it Hot Carriers in Semiconductor 
Nanostructures: Physics and Applications}, edited by J. Shah (Academic, San
Diego 1992), p. 191.
\bibitem{Rossi92a}
F. Rossi, P. Poli, and C. Jacoboni, 
Semicond. Sci. Technol. {\bf 7}, 1017 (1992).
\bibitem{Mysyrowicz86}
A. Mysyrowicz, D. Hulin, A. Antonetti, A. Migus, W.T. Masselink, and
H. Morko\c{c}, 
Phys. Rev. Lett. {\bf 56}, 2748 (1986).
\bibitem{Peyghambarian89}
N. Peyghambarian, S.W. Koch, M. Lindberg, B.D. Fl\"ugel, M. Joffre, 
Phys. Rev. Lett. {\bf 62}, 1185 (1989).
\bibitem{Knox89}
W.H. Knox, D.S. Chemla, D.A.B. Miller, J.B. Stark, and S. Schmitt-Rink,
Phys. Rev. Lett. {\bf 1989}, 1189 (1989).
\bibitem{Cundiff94}
S.T. Kundiff, A. Knorr, J. Feldmann, S.W. Koch, E.O. G\"obel, H. Nichel, 
Phys. Rev. Lett. {\bf 73}, 1178 (1994).
\bibitem{Becker88}
P.C. Becker, H.L. Fragnito, C.H. Brito Cruz, R.L. Fork, J.E. Cunningham,
J.E. Henry, and C.V. Shank, 
Phys. Rev. Lett. {\bf 61}, 1647 (1988).
\bibitem{Masumoto83}
Y. Masumoto, S. Shionoya, and T. Takagahara, 
Phys. Rev. Lett. {\bf 51}, 923 (1983).
\bibitem{Schultheis86}
L. Schultheis, J. Kuhl, A. Honold, and C.W. Tu, 
Phys. Rev. Lett. {\bf 57}, 1635 (1986); Phys. Rev. Lett. {\bf 57}, 1797 (1986). 
\bibitem{Noll90}
G. Noll, U. Siegner, S.G. Shevel, and E.O. G\"obel, 
Phys. Rev. Lett. {\bf 64}, 792 (1990).
\bibitem{Langer90}
V. Langer, H. Stolz, and W. von~der Osten, 
Phys. Rev. Lett. {\bf 64}, 854 (1990).
\bibitem{Gobel90}
E.O. G\"obel, K. Leo, T.C. Damen, J. Shah, S. Schmitt-Rink, W. Sch\"afer,
J.F. M\"uller, and K. K\"ohler, 
Phys. Rev. Lett. {\bf 64}, 1801 (1990).
\bibitem{Frohlich91}
D. Fr\"ohlich, A. Kulik, B. Uebbing, A. Mysyrowicz, V. Langer, H. Stolz, 
and
W. von~der Osten, Phys. Rev. Lett. {\bf 67}, 2343 (1991).
\bibitem{Stolz91}
H. Stolz, V. Langer, E. Schreiber, S.A. Permogorov, W. von der Osten, 
Phys. Rev. Lett. {\bf 67}, 679 (1991).
\bibitem{Leo91}
K. Leo, J. Shah, E.O. G\"obel, T.C. Damen, S. Schmitt-Rink, W. Sch\"afer, 
and K. K\"ohler, Phys. Rev. Lett. {\bf 66}, 201 (1991).
\bibitem{Feldmann93}
J. Feldmann, T. Meier, G. von Plessen, M. Koch, E.O. G\"obel, P. Thomas, G.
Bacher, C. Hartmann, H. Schweizer, W. Sch\"afer, and H. Nickel,
Phys. Rev. Lett. {\bf 70}, 3027 (1993).
\bibitem{Waschke93}
C. Waschke, H.G. Roskos, R. Schwedler, K. Leo, H. Kurz, and K. K\"ohler,
Phys. Rev. Lett. {\bf 70}, 3319 (1993).
\bibitem{Leo90}
K. Leo, M. Wegener, J. Shah, D.S. Chemla, E.O. G\"obel, T.C. Damen,
S. Schmitt-Rink, and W. Sch\"afer, 
Phys. Rev. Lett. {\bf 65}, 1340 (1990).
\bibitem{Weiss92}
S. Weiss, M.-A. Mycek, J.-Y. Bigot, S. Schmitt-Rink, and D.S. Chemla,
Phys. Rev. Lett. {\bf 69}, 2685 (1992).
\bibitem{Kim92a}
D.-S. Kim, J. Shah, J.E. Cunningham, T.C. Damen, W. Sch\"afer, M. Hartmann,
and S. Schmitt-Rink, 
Phys. Rev. Lett. {\bf 68}, 1006 (1992).
\bibitem{Kim92b}
D.-S. Kim, J. Shah, T.C. Damen, W. Sch\"afer, F. Jahnke, S. Schmitt-Rink, 
and K. K\"ohler, 
Phys. Rev. Lett. {\bf 69}, 2725 (1992).
\bibitem{Roskos92}
H.G. Roskos, M.C. Nuss, J. Shah, K. Leo, D.A.B. Miller, A.M. Fox,
S. Schmitt-Rink, and K. K\"ohler, 
Phys. Rev. Lett. {\bf 68}, 2216 (1992).
\bibitem{Planken92}
P.C.M. Planken, M.C. Nuss, I. Brener, K.W. Goossen, M.S.C. Luo, S.L. Chuang, 
and L. Pfeiffer, Phys. Rev. Lett. {\bf 69}, 3800 (1992).
\bibitem{Comte86}
C. Comte and G. Mahler, Phys. Rev.B {\bf 34}, 7164 (1986).
\bibitem{Schmitt86a}
S. Schmitt-Rink, C. Ell, and H. Haug, 
Phys. Rev. B{\bf 33}, 1183 (1986).
\bibitem{Schmitt88}
S. Schmitt-Rink, D.S. Chemla, and H. Haug, Phys. Rev. B{\bf 37}, 941 (1988).
\bibitem{Henneberger88}
K. Henneberger and H. Haug, Phys. Rev. B{\bf 38}, 9759 (1988).
\bibitem{Kuznetsov91}
A.V. Kuznetsov, 
Phys. Rev. B{\bf 44}, 8721 (1991).
\bibitem{Haug92}
H. Haug and C. Ell, Phys. Rev. B{\bf 46}, 2126 (1992).
\bibitem{Balslev89}
I. Balslev, R. Zimmermann, and A. Stahl, Phys. Rev. B{\bf 40}, 4095 (1989).
\bibitem{Schmitt86b}
S. Schmitt-Rink and D.S. Chemla, Phys. Rev. Lett. {\bf 57}, 2752 (1986).
\bibitem{Lindberg88}
M. Lindberg and S.W. Koch, Phys. Rev. B{\bf 38}, 3342 (1988).
\bibitem{Wegener90}
M. Wegener, D.S. Chemla, S. Schmitt-Rink, and W. Sch\"afer, 
Phys. Rev. A{\bf 42}, 5675 (1990).
\bibitem{Schafer93}
W. Sch\"afer, F. Jahnke, and S. Schmitt-Rink, 
Phys. Rev. B{\bf 47}, 1217 (1993).
\bibitem{Binder94}
E. Binder, T. Kuhn, and G. Mahler, Phys. Rev. {\bf B50}, 18319 (1994).
\bibitem{Hess96}
O.~Hess and T.~Kuhn, Phys. Rev. A{ \bf 54}, 3347 (1996);
ibid. 3360 (1996).
\bibitem{Kuhn97}
T. Kuhn, in {\it Theory of Transport Properties of
  Semiconductor Nanostructures}, edited by E. Sch\"oll 
(Chapman \& Hall, London 1997).
\bibitem{Fork87}
R.L. Fork, C.H. Brito Cruz, P.C. Becker, and C.V. Shank, 
Phys. Rev. Lett. {\bf 12}, 483 (1987).
\bibitem{Kuhn92a}
T. Kuhn and F. Rossi, Phys. Rev. Lett. {\bf 69},  977  (1992).
\bibitem{Kuhn92b}
T. Kuhn and F. Rossi, Phys. Rev. B {\bf 46},  7496  (1992).
\bibitem{Rossi93}
F. Rossi, S. Haas, and T. Kuhn, 
Semicond. Sci. Technol. {\bf 9}, 411 (1994).
\bibitem{Rossi94}
F. Rossi, S. Haas, and T. Kuhn, Phys. Rev. Lett. {\bf 72}, 152 (1994).
\bibitem{Haas96}
S. Haas, F. Rossi, and T. Kuhn, Phys. Rev. B {\bf 53},  12855  (1996).
\bibitem{Lohner93}
A. Lohner, K. Rick, P. Leisching, A. Leitenstorfer, T. Elsaesser, T. Kuhn, 
F. Rossi, and W. Stolz, Phys. Rev. Lett. {\bf 71},  77  (1993).
\bibitem{Leitenstorfer94a}
A. Leitenstorfer, A. Lohner, K. Rick, P. Leisching, T. Elsaesser, T. Kuhn, 
F. Rossi, W. Stolz, and K. Ploog, Phys. Rev. B {\bf 49},  16372  (1994).
\bibitem{Leitenstorfer94b}
A. Leitenstorfer, A. Lohner, T. Elsaesser, S. Haas, F. Rossi, T. Kuhn, W. 
Klein, G. Boehm, G. Traenkle, and G. Weimann, 
Phys. Rev. Lett. {\bf 73},  1687  (1994).
\bibitem{Kuhn95}
T. Kuhn, F. Rossi, A. Leitenstorfer, A. Lohner, T. Elsaesser, W. Klein, 
G. Boehm, G. Traenkle, and G.W. Weimann, 
in {\it Hot Carriers in Semiconductors}, edited by K.
  Hess, J.-P. Leburton, and U. Ravaioli (Plenum Press, New York, 1996), 
p.\ 199.
\bibitem{Leitenstorfer96b}
A. Leitenstorfer, T. Elsaesser, F. Rossi, T. Kuhn, W. Klein, G. Boehm, G. 
Traenkle, and G.W. Weimann, Phys. Rev. B {\bf 53},  9876  (1996).
\bibitem{Bloch46}
F. Bloch, Phys. Rev. {\bf 70}, 460 (1946). 
\bibitem{Kash89} 
J.A. Kash and J.C. Tsang, 
in {\it{Light Scattering in Solids VI}}, 
edited by M.~Cardona and G.G\"untherodt, Springer, Berlin (1989), p.~423.
\bibitem{Poetz83} 
W. P\"otz and P. Kocevar, Phys. Rev. B {\bf 28}, 7040 (1983).
\bibitem{Quade94}
W. Quade, E. Sch\"oll, F. Rossi, and C. Jacoboni, Phys. Rev. B {\bf 50},  7398
  (1994).
\bibitem{Rossi92b}
F. Rossi, R. Brunetti, and C. Jacoboni,  in {\it {H}ot {C}arriers in
  {S}emiconductor {N}anostructures: {P}hysics and {A}pplications}, edited by J.
  Shah (Academic Press inc., Boston, 1992), p.\ 153.
\bibitem{Brunetti89}
R. Brunetti, C. Jacoboni, and F. Rossi, Phys. Rev. B {\bf 39},  10781  (1989).
\bibitem{TranThoai93}
D.B. Tran Thoai and H. Haug, Phys. Rev. B {\bf 47}, 3574 (1993).
\bibitem{Sayed94}
K.E. Sayed, L. B\`anyai, and H. Haug, Phys. Rev. {\bf B50}, 1541 (1994).
\bibitem{Schilp94}
J. Schilp, T. Kuhn, and G. Mahler, Phys. Rev. B {\bf 50}, 5435 (1994).
\bibitem{Meden95}
V. Meden, C. W\"ohler, j. Fricke, and K. Sch\"onhammer, 
Phys. Rev. {\bf B52}, 5624 (1995).
\bibitem{Kane94}
M.G. Kane, K.W. Sun, and S.A. Lyon, Semicond. Sci. Technol. {\bf 9}, 697 
(1994).
\bibitem{Kash95}
J.A. Kash, Phys. Rev. {\bf B51}, 4680 (1995).
\bibitem{Bloch28}
F. Bloch, Z. Phys. {\bf 52}, 555 (1928).
\bibitem{Zener34}
C. Zener, Proc. R. Soc. {\bf 145}, 523 (1934).
\bibitem{Houston40}
W.V. Houston, Phys. Rev. {\bf 57}, 184 (1940).
\bibitem{Kane59}
E.O. Kane, J. Phys. Chem. Solids {\bf 12}, 181 (1959)
\bibitem{Argyres62}
P.N. Argyres, Phys. Rev. {\bf 126}, 1386 (1962).
\bibitem{Wannier60}
G.H. Wannier, Phys. Rev. {\bf 117}, 432 (1960).
\bibitem{Zak68a}
J. Zak, Phys. Rev. Lett. {\bf 20}, 1477 (1968).
\bibitem{Kittel63}
C. Kittel, {\it Quantum Theory of Solids} (Wiley, New York, 1963), p. 190.
\bibitem{Krieger86}
J.B. Krieger and G.J. Iafrate, Phys. Rev. B {\bf 33}, 5494 (1986).
\bibitem{Nenciu91}
G. Nenciu, Rev. Mod. Phys. {\bf 63}, 91 (1991).
\bibitem{Bastard89}
G. Bastard, {\it Wave Mechanics of Semiconductor Heterostructures}, Les 
Editions de Physique (Les Ulis, France, 1989).
\bibitem{Mendez88}
E.E. Mendez, F. Agullo-Rueda, and J.M. Hong, Phys. Rev. Lett. {\bf 60}, 
2426 (1988).
\bibitem{Voisin88}
P. Voisin, J. Bleuse, C. Bouche, S. Gaillard, C. Alibert, and A, Regreny, 
Phys. Rev. Lett. {\bf 61}, 1639 (1988).
\bibitem{Feldmann92}
J. Feldmann, K. Leo, J. Shah, D.A.B. Miller, J.E. Cunningham, T. Meier, G. 
von Plessen, A. Schulze, P. Thomas, and S. Schmitt-Rink, 
Phys. Rev. B {\bf 46}, 7252 (1992).
\bibitem{vonPlessen92}
G. von Plessen and P. Thomas, Phys. Rev. B {\bf 45}, 9185 (1992).
\bibitem{Leo92}
K. Leo, P.H. Bolivar, F. Br\"uggemann, R. Schwedler, Solid State Commun. 
{\bf 84}, 943 (1992).
\bibitem{Leisching94}
P. Leisching, P. Haring Bolivar, W. Beck, Y. Dhaibi, F. Bruggemann, 
R. Schwedler, H. Kurz, K. Leo, and K. K\"ohler,
Phys. Rev. B {\bf 50}, 14389 (1994).
\bibitem{Dekorsy94}
T. Dekorsy, P. Leisching, K. Kohler, and H. Kurz, Phys. Rev. B {\bf 50}, 8106 
(1994).
\bibitem{Dekorsy95}
T. Dekorsy, R. Ott, H. Kurz, and K. Kohler, Phys. Rev. B {\bf 51}, 17275 
(1995).
\bibitem{Roskos94}
H.G. Roskos, C. Waschke, R. Schwedler, P. Leisching, Y. Dhaibi, H. Kurz, 
and K. K\"ohler, Superlattices and Microstructures {\bf 15}, 281 (1994).
\bibitem{Rossi97}
F. Rossi, in {\it Theory of Transport Properties of
  Semiconductor Nanostructures}, edited by E. Sch\"oll 
(Chapman \& Hall, London 1997).
\bibitem{Rossi95a}
F. Rossi, T. Meier, P. Thomas, S.W. Koch, P.E. Selbmann, and E. Molinari, 
Phys. Rev. B {\bf 51}, 16943 (1995).
\bibitem{Rossi95b}
F. Rossi, T. Meier, P. Thomas, S.W. Koch, P.E. Selbmann, and E. Molinari, 
in {\it Hot Carriers in Semiconductors}, edited by K.
  Hess, J.-P. Leburton, and U. Ravaioli (Plenum Press, New York, 1996), 
p.\ 157.
\bibitem{Meier95b}
T. Meier, F. Rossi, P. Thomas, and S.W. Koch, Phys. Rev. Lett. {\bf 75},  2558
  (1995).
\bibitem{Je95}
K.-C. Je, T. Meier, F. Rossi, and S.W. Koch, Appl. Phys. Lett. {\bf 67},  2978
  (1995).
\bibitem{Koch95}
S.W. Koch, T. Meier, T. Stroucken, A. Knorr, J. Hader, F. Rossi, and P. 
Thomas, in {\it Microscopic theory of semiconductors: quantum
  kinetics, confinement and lasers}, edited by S.W. Koch (World Scientific,
  Singapore, 1995), p.\ 81.
\bibitem{Rossi96a}
F. Rossi, M. Gulia, P.E. Selbmann, E. Molinari, T. Meier, P. Thomas, and 
S.W. Koch, in {\it Proc. 23rd {ICPS}, Berlin, Germany}, edited by
M. Scheffler and R. Zimmermann (World Scientific, Singapore, 1996), p.\ 1775.
\bibitem{Ruecker92} 
H. R\"ucker, E. Molinari, and P. Lugli,
Phys. Rev. B {\bf 45}, 6747 (1992).
\bibitem{Molinari94}
E. Molinari, in {\it{Confined Electrons and Photons: New Physics
and Applications}},
edited by E. Burstein and C. Weisbuch (Plenum, New York, 1994).
\bibitem{vonPlessen94}
G. von Plessen, T. Meier, J. Feldmann, E.O. G\"obel, P. Thomas, K.W. 
Goossen, J.M. Kuo, and R.F. Kopf, Phys. Rev. B {\bf 49}, 14058 (1994).
\bibitem{Dignam90}
M.M. Dignam and J.E. Sipe,
Phys. Rev. Lett. {\bf 64}, 1797 (1990).
\bibitem{review_excitons} 
See e.g. {\it Proc. 4th Internat. Conf. on Optics of Excitons in Confined
Systems}, Il Nuovo Cimento D vol. 17 (1995).
\bibitem{review_wires} 
For a review see:
R.~Cingolani and R.~Rinaldi, Rivista Nuovo Cimento {\bf 16}, 1 (1993).
\bibitem{Ogawa91a} 
T. Ogawa and T. Takagahara, 
Phys. Rev. {\bf B43}, 14325 (1991).
\bibitem{Ogawa91b} 
T. Ogawa and T. Takagahara, Phys. Rev. {\bf B44}, 8138 (1991).
\bibitem{Rossi96b}
F. Rossi and E. Molinari, Phys. Rev. Lett. {\bf 76}, 3642 (1996).
\bibitem{Rossi96c}
F. Rossi and E. Molinari, Phys. Rev. {\bf B53}, 16462 (1996).
\bibitem{Rossi96d}
F. Rossi and E. Molinari, 
in {\it Proc. 23rd {ICPS}, Berlin, Germany}, edited by
M. Scheffler and R. Zimmermann (World Scientific, Singapore, 1996), p.\ 1161.
\bibitem{Rinaldi94} 
R.~Rinaldi, R. Cingolani, M. Lepore, M. Ferrara, I.M. Catalano, F. Rossi, 
L. Rota, E. Molinari, P. Lugli, U. Marti, D. Martin, F. Morrier-Gemoud, P. 
Ruterana, and F.K. Reinhart,
Phys. Rev. Lett. {\bf 73}, 2899 (1994).
\bibitem{Sakaki96} 
T. Someya, H. Akiyama, and H. Sakaki,
Phys. Rev. Lett. {\bf 76}, 2965 (1996).

\end{references}
\end{document}